\documentclass[%
 reprint,
superscriptaddress,
 amsmath,amssymb,
 aps,
]{revtex4-2}

\usepackage{graphicx}
\usepackage{dcolumn}
\usepackage{bm}
\usepackage[figuresright]{rotating}
\usepackage{multirow}
\usepackage{xcolor}
\usepackage{hyperref}
\hypersetup{
    colorlinks=true,
    linkcolor=blue,
    citecolor=blue,      
    urlcolor=blue}
\newcommand\apjl{Astrophys. J. Lett.} % Astrophysical Journal, Letters 
\newcommand\mnras{ Mon. Not. R. Astron. Soc.} % Monthly Notices of the RAS 
\newcommand\na{New Astron.} % New Astronomy

\begin{document}

\preprint{APS/123-QED}

\title{Prospects for reconstructing the gravitational-wave signals from core-collapse supernovae with Advanced LIGO-Virgo and the BayesWave algorithm}

\author{Nayyer Raza}
 \email{nayyer.raza@mail.mcgill.ca}
 \affiliation{Department of Physics and Astronomy, University of British Columbia, Vancouver, BC V6T1Z1, Canada}
 \affiliation{Department of Physics, McGill University, Montr\'{e}al, QC H3A2T8, Canada}
\author{Jess McIver}
 \email{mciver@phas.ubc.ca}
 \affiliation{Department of Physics and Astronomy, University of British Columbia, Vancouver, BC V6T1Z1, Canada}
\author{Gergely D\'{a}lya}
 \affiliation{Institute of Physics, E\"{o}tv\"{o}s University, 1117 Budapest, Hungary}
 \affiliation{Department of Physics and Astronomy, Universiteit Gent, B-9000 Gent, Belgium}
\author{Peter Raffai}
 \affiliation{Institute of Physics, E\"{o}tv\"{o}s University, 1117 Budapest, Hungary}

\date{\today}

\begin{abstract}

Our current understanding of the core-collapse supernova explosion mechanism is incomplete, with multiple viable models for how the initial shock wave might be energized enough to lead to a successful explosion. Detection of a gravitational-wave signal emitted in the initial few seconds after stellar core-collapse would provide unique and crucial insight into this process. With the Advanced LIGO and Advanced Virgo detectors expected to approach their design sensitivities soon, we could potentially detect this signal from a supernova within our galaxy. In anticipation of such a scenario, we study how well the BayesWave algorithm can recover the gravitational-wave signal from core-collapse supernova models in simulated advanced detector noise, and optimize its ability to accurately reconstruct the signal waveforms. We find that BayesWave can confidently reconstruct the signal from a range of supernova explosion models in Advanced LIGO-Virgo for network signal-to-noise ratios $\gtrsim 30$, reaching maximum reconstruction accuracies of $\sim 90\%$ at SNR $\sim 100$. For low SNR signals that are not confidently recovered, our optimization efforts result in gains in reconstruction accuracy of up to $20-40\%$, with typical gains of $\sim 10\%$.

\end{abstract}

\maketitle

\section{Introduction} \label{sec:intro}

Despite many detailed multi-wavelength electromagnetic observations of core-collapse supernovae (CCSNe), the exact mechanism powering these explosions is not yet fully understood (see e.g. \citet{ Burrows2021} and references therein). This is in part because the electromagnetic signal only escapes from the very outer layers of the star after shock break-out (SBO), on time scales of hours to days after the core collapses \citep{Kistler2013}. However, most of the critical explosion physics occurs in the very central core of the star within the first few seconds \citep{Janka2012}. The thermodynamic state of the explosion in this crucial period is imprinted on to the escaping neutrino flux, which carries away almost all of the total explosion energy, $\sim 10^{53} ~\mathrm{ergs}$ \citep{Muller2019}. Non-spherical, accelerated mass motions in the dense supernova core at this time also produce gravitational waves (GWs), lasting $\lesssim 1 $ s after core bounce, which then probe the internal dynamics of the explosion mechanism (see e.g. \citet{Abdikamalov2020arXiv} for a recent review). We can thus use multi-messenger astronomy to study these highly energetic events and help solve the mystery of what powers CCSNe.

While Supernova SN1987A marked the first (and so far only) supernova for which the neutrino signal was also observed \citep{Hirata1987, Bionta1987, Alexeyev1988}, GWs from CCSNe have not yet been detected \citep{LVC2020, LVK2021arXiv}. This is especially challenging since CCSNe are rare events, expected to only occur approximately once or twice per century in large galaxies such as the Milky Way \citep{Rozwadowska2021}.

The Advanced LIGO \citep{aLIGO2015} and Advanced Virgo \citep{AdVirgo2015} ground-based interferometric GW detectors are expected to achieve their design sensitivities for their fourth observing run \citep{LVK2020}. At these sensitivities (along with the addition of new detectors such as KAGRA \citep{Somiya2012,Aso2013} and LIGO-India \citep{Iyer2011} in the future), we could potentially observe a Galactic CCSN with GWs \citep{Gossan2016, Szczepanczyk2021}. In anticipation of such an event, in this study we follow up on the CCSN sensitivity analyzed in the all-sky short duration burst search of the most recent LIGO-Virgo observing run \citep{LVK2021arXiv}, and characterize how well the BayesWave algorithm \citep{Cornish2015} can reconstruct CCSN models embedded in Advanced LIGO and Advanced Virgo design sensitivity detector noise. We also aim to tune BayesWave for CCSN signals, and produce a set of recommendations to be used for analysis on a candidate detection that will optimize the reconstruction accuracy.

\subsection{Core-collapse supernovae}

CCSNe are the end stage of the life of massive stars with $M \gtrsim 8 ~\mathrm{M_{\odot}}$. When the mass of the central iron core exceeds the effective Chandrasekhar mass, runaway collapse of the core begins, marking the onset of the explosion. This value can vary between $M_{Ch} \sim 1.34-1.8~\mathrm{M_{\odot}}$, depending on a range of different properties of the progenitor star (see e.g. \citet{Woosley2002} for a classic review). A shock forms between the supersonically infalling inner core of $\sim 0.6-0.8 ~\mathrm{M_{\odot}}$ and the subsonically infalling outer core of $\sim 0.6 ~\mathrm{M_{\odot}}$. As the temperature and density increases, the collapse is accelerated by photo-dissociation and electron capture, which also produces a burst of neutrinos that are initially trapped in the ultra-dense inner core. As the central density reaches nuclear density, the inner core cannot collapse any further and rebounds, transferring in-falling energy-momentum by launching a shock wave outwards.

This shock wave loses significant energy as it propagates through the still in-falling outer core and heats it, dissociating iron nuclei into free protons and neutrons. Furthermore, as the density behind the shock gets low enough, $\sim 10^{11} ~\mathrm{g\cdot cm^{-3}}$, the neutrinos start to escape, contributing to further energy loss. Eventually, the shock stalls, around 100-200 km from the center (a few tens of milliseconds after bounce), near the outer iron layer. From this point, to get a successful explosion the shock must be revived within a few hundred milliseconds before the continued accretion onto the central proto neutron star (PNS) triggers further collapse into a black hole.

The exact mechanism behind the re-energization of the shock wave and the subsequent successful explosion is not fully understood \citep{Burrows2021}. The delayed neutrino heating mechanism has emerged as a promising candidate to explain the shock revival for most slowly rotating progenitors. In this scenario the intense neutrino flux from the hot PNS deposits a small fraction of its energy ($\sim 1\%$) into the gain region behind the stalled shock (see e.g. \citet{Janka2017} for a review of neutrino-driven explosions). For rapidly rotating stars with strong magnetic fields (expected to account for $\sim 1\%$ of CCSN progentiors \citep{Woosley2006}), magnetohydrodynamic effects can allow the transfer of energy from the highly magnetized PNS into the outer stellar layers for a violent explosion, in the magnetorotationally-driven mechanism (e.g. \cite{Obergaulinger2020}). While it's expected that neutrino heating plays a key role in most explosion models, the exact role of the different potential drivers of the explosion, which include the neutrino flux, differential rotation of the progenitor star, and strong magnetic fields, remains to be fully understood. We direct the reader to the review by \citet{Janka2012} for a detailed discussion of the different possible CCSN mechanisms, and to the recent review by \citet{Burrows2021} and the references therein for current limitations in our understanding.

\subsection{Gravitational waves from CCSNe}

The GW signal can provide unique insight into the state of the PNS in this crucial time period, and inform about the explosion mechanism \citep{Abdikamalov2020arXiv}. Some expected features in a GW signal are common to most CCSN models. A quadrupole moment from the deformation of the PNS is expected to occur after the core bounce. Waves from the outer convective region are likely to travel in and strike the outer core, and waves from convection in the PNS are likely to internally excite core and surface oscillations. For slowly rotating stars, the GW signal is then expected to be dominated by the fundamental oscillation modes of the PNS (f, g, p modes), which increase in frequency with time as the PNS contracts (e.g. \cite{Torres-Forne2018, OConnor2018, Radice2019, Powell2019}). The exact shape and amplitude of the signal can then also place constraints on the source properties such as the PNS mass, size, core compactness and explosion energy (e.g. \cite{Torres-Forne2019, Bizouard2021, Sotani2021}).

Bulk fluid motion from the advective acoustic cycle, particularly in the gain region, may also cause global asymmetric perturbations leading to emission in the low-frequency regime. Linear or spiral oscillation modes (l, m) of the shock front may arise from the Standing Accretion Shock Instability (SASI) \citep{Blondin2003}. %Low-frequency features ($\sim$ 100-200 Hz) may also suggest a soft nuclear matter equation of state.

In the case that the explosion is powered by strong magnetic fields or amplified by a high rotation rate, prompt broadband emission can be expected within $\sim 20$ ms after core bounce (e.g. \cite{Takiwaki2011,Richers2017}). A rapidly rotating stellar core is deformed due to its angular momentum and leads to the derivatives of the quadrupole moment to change significantly as the core collapses, and the core bounce prompt convection signal can become a dominant feature (e.g. \cite{Abdikamalov2014,Fuller2015})

\subsection{Detecting the GW signal with Advanced LIGO-Virgo}

The stochastic nature of the CCSN explosion makes predicting exactly which features will occur currently impossible. However, the dominant emission features that are expected from the GW signal of different CCSN models, $\sim$ 20 Hz - 2 kHz, lie in the most sensitive frequency regime of the Advanced LIGO and Virgo detectors \cite{LVK2020}. This sensitivity overlap makes CCSN within our Galaxy a potentially promising target for current detectors. While the expected event rate of 1-2 per century \citep{Rozwadowska2021} for such events is low, a detection of the associated GW signal should a Galactic CCSN happen within the next observing run would be invaluable, and could allow us to distinguish between different explosion models and mechanisms (e.g. \cite{Logue2012,Roma2019}), as well as extract astrophysical parameters of the source progenitor and remnant.

GW detection and analysis algorithms have been developed over the years targeted towards characterizing short duration GW signals in the LIGO and Virgo detectors, and some of these have been used to study GWs from CCSN. In particular, previous work by \citet{Gossan2016} used the X-Pipeline algorithm \citep{Sutton2010} and GW waveform models from multi-dimensional simulations to study in detail the detectability of these models for the advanced era detectors. The recent study by \citet{Szczepanczyk2021} used the coherent WaveBurst algorithm \citep{Klimenko2016} to analyze it's ability to detect and reconstruct CCSN GW signals, with an expanded and updated set of waveform models, for the anticipated fourth and fifth observing runs.

Another candidate analysis algorithm is BayesWave, which is designed to reconstruct the signal from GW burst sources (short duration transient signals) using a wavelet basis, while making minimal assumptions about the specific waveform morphology \cite{Cornish2015}. This un-modeled approach, as opposed to matched filtering using known waveforms for compact binary coalescence candidates (e.g. \cite{Adams2016,Usman2016,Messick2017}), makes BayesWave a suitable waveform reconstruction pipeline for CCSN studies.

To prepare for a candidate detection, we aim to study and quantify BayesWave's potential ability to reconstruct different types of CCSN signals. Furthermore we aim to identify which algorithm settings are most sensitive to CCSN signals, and produce a set of recommendations to be used for analysis on a real signal. In order to achieve these goals of characterizing BayesWave's performance, we study five different recent CCSN explosion models from three dimensional simulations. These models encompass the same set used in \citet{LVK2021arXiv}, and span a range of explosion models, including failed explosions, neutrino driven explosions and magnetorotationally driven explosions.

The rest of the paper is organized as follows. In Section \ref{sec:methods} we describe the BayesWave algorithm, the supernova waveform models we employed, the general procedure for our reconstruction studies, and our optimization efforts. In Section \ref{sec:results} we present our results, highlighting BayesWave's reconstruction accuracy for CCSN, gains from our optimization, and the recovery of a realistic distribution of sources in the Milky Way. Section \ref{sec:discuss} follows with a discussion of BayesWave's performance between the different models, the distances to which it can reconstruct CCSN signals, and the limitations of BayesWave in this work. We summarize and conclude with final remarks and future work in Section \ref{sec:conclusion}.

\section{Methods} \label{sec:methods}

The aim of our study is to analyze and optimize the reconstruction of CCSN GW signals with BayesWave. In the next few subsections we detail the methods that we employ, including a summary of the BayesWave algorithm and its features, a description of the CCSN waveform models selected, the steps taken to perform the reconstructions in simulated Advanced LIGO-Virgo detectors, and the optimization efforts employed in the procedure to yield more reliable and accurate reconstructions.

\subsection{BayesWave}
    
BayesWave is a GW signal reconstruction algorithm for short bursts that makes minimal assumptions about the signal morphology \citep{Cornish2015}. For each detector BayesWave models the analyzed data ($d$) as a linear combination of the GW signal in the detector frame ($s$), the detector's stationary Gaussian noise ($n$), and transient instrumental noise (glitches, $g$): $d = s + n + g$. The ratio of the Bayesian evidences for each model determines which has the most support.

In practice BayesWave uses a trans-dimensional reversible jump Markov chain Monte Carlo (RJMCMC) to place a variable number of Morlet-Gabor (sine-Gaussian) wavelets, the linear combination of which forms a reconstructed signal. For a multiple detector configuration, the wavelets placed for the signal model must be coherent across detectors, requiring the same set of extrinsic parameters (sky location, ellipticity, polarization angle) to properly project the signal onto the detectors. For the glitch model, however, the wavelets are uncorrelated.
      
Since the supernova analysis pipeline must make minimal assumptions about the waveform, BayesWave is then one of the more suitable choices to use in our study. Because it is agnostic to the type or source of the signal, BayesWave has been extensively used for un-modeled reconstructions of GW signals from compact binary coalescences (e.g. \cite{Chatziioannou2017,LVC2019,Ghonge2020,Dalya2021}) and white-noise bursts \citep{Becsy2017}, as well as a glitch subtraction tool to `clean' a portion of the detector data that has an instrumental noise event overlapping with a real signal (e.g. \cite{Pankow2018, Chatziioannou2021}). It could also potentially be used as a follow-up tool to cWB to help better discriminate CCSNe events from instrumental noise events \citep{Gill2018}.

If the analysis involves recovering a known signal that has been injected (added) into detector noise (as is the case for our study), then BayesWave also calculates the overlap (also called match) between the waveform of the injected signal $h_i$ and the BayesWave wavelet reconstruction $h_r$. The overlap is defined as

\begin{equation}\label{eq:overlap}
O = \frac{(h_i|h_r)}{\sqrt{(h_i|h_i)(h_r|h_r)}},
\end{equation}

where the notation $(x|y)$ denotes a noise-weighted inner product using the detector's one-sided noise power spectral density $S_h(f)$:

\begin{equation}\label{eq:inner_product}
(x|y) = 2 \int_{0}^{\infty} \frac{x(f)y^{*}(f) + x^{*}(f)y(f)}{S_h(f)} df
\end{equation}

Thus the overlap characterizes the accuracy of the waveform reconstruction, measuring the similarity of the injected and recovered waveforms. This value ranges from -1 to 1, with $O = 1$ meaning a perfect match between the two waveforms, $O = 0$ meaning no match, and $O = -1$ indicating a perfect anti-correlation. In our work we quote the combined weighted network overlap for all $N=3$ detectors, which is computed as 

\begin{equation}\label{eq:overlap_net}
O_{net} = \frac{\sum_{j=1}^N (h_i^{(j)}|h_r^{(j)})}{\sqrt{\sum_{j=1}^N (h_i^{(j)}|h_i^{(j)}) \sum_{j=1}^N (h_r^{(j)}|h_r^{(j)})}},
\end{equation}

where $j$ denotes the $j$-th detector.

In the context of our study, this means that if the BayesWave reconstructed CCSN waveform has an overlap of $O = 0.5$, for example, then that reconstructed waveform accurately matches 50\% of the total time-frequency and phase signature of the true waveform. In the rest of this paper, we then often say that the reconstruction accuracy is $X\%$ to reflect the fact that the overlap between the reconstructed and injected signals is $X/100$. Furthermore, as outlined in Section \ref{subsec:gen_proc}, in our studies BayesWave produces 20000 reconstructed waveforms for each injected signal, and calculates the overlap for each of these. The result is a distribution of overlap values, from which we select the median value as the characteristic overlap, and compute the standard deviation of the distribution as an estimate of the overlap uncertainty.

\subsection{Supernova waveform models}
Here we briefly describe the supernova waveform models from 3D simulations used in our study and the GW features they exhibit (as shown in the first row in Fig. \ref{fig:spectrograms}). We refer the interested reader to their source papers for more detailed information about the details of their simulation procedure and modeling techniques.
\begin{itemize}
    \item The \citet{OConnor2018} \textit{mesa20-3D-pert} model (hereafter referred to as \textit{m20-3Dp}) has a $\mathrm{20 \ M_{\odot}}$ solar metallicity non-rotating progenitor star, and fails to explode within $\sim 500$ ms after core bounce, the extent of the simulation. However, the model does emit a GW signal that is dominated by the g-mode oscillation of the PNS surface, which begins at $\sim 300$ Hz and grows with time, peaking at $\sim 600-700$ Hz at $\sim 200$ ms after core bounce. A subdominant signal due to the spiral SASI modes is also present at $\sim 50-200$ Hz.
    \item The \citet{Radice2019} \textit{s9} model has a $\mathrm{9 \ M_{\odot}}$ solar metallicity non-rotating progenitor, and represents a lower mass neutrino-driven explosion. A broadband prompt convection signal develops $\sim 50$ ms after bounce in conjunction with the neutrino shock breakout. After a short quiescent period, the PNS g-mode oscillation begins, rising to frequencies of $\sim 700$ Hz. The shock is successfully revived at $\sim 200$ ms after core bounce, and no significant GW emission occurs after $\sim 300$ ms.
    \item The \citet{Powell2019} \textit{s18} model has a $\mathrm{18 \ M_{\odot}}$ solar metallicity non-rotating progenitor, and represents a typical neutrino driven explosion. GW emission from g-mode surface oscillations of the PNS begins $\sim 100$ ms after core bounce, and peaks at $\sim 800-1000$ Hz just after shock revival and onset of the explosion $\sim 300$ ms after bounce. Significant GW emission continues up to $\sim 600$ ms.
    \item The \citet{Powell2020} \textit{m39} model has a rapidly rotating Wolf-Rayet star as its progenitor with a $\mathrm{39 \ M_{\odot}}$ helium star mass, 2\% solar metallicity, and an initial surface rotation velocity of $600 ~\mathrm{km \cdot s^{-1}}$. It is a strong neutrino driven explosion, with GW features amplified by the rapid rotation. Strong GW emission from prompt convection at low frequencies occurs shortly after bounce time, mainly visible in observer angles towards the equator. There is also a strong core bounce signal at frequencies $> 500$ Hz. This is followed by emission associated with f-mode oscillations of the PNS, which peaks at a frequency of $\sim 750$ Hz about 400 ms after core bounce. The shock revival occurs at $\sim 200$ ms, and GW emission continues up to $\sim 600$ ms.
    \item The \citet{Obergaulinger2020} \textit{35OC-RO} model has a $\mathrm{35 \ M_{\odot}}$ rapidly rotating progenitor with strong magnetic fields, with a sub-solar metallicity and an equatorial surface rotation velocity of $380 ~\mathrm{km \cdot s^{-1}}$. It represents a strong magneto-rotationally driven explosion. A bounce signal is quickly followed by an evolving PNS oscillation track. After the shock is revived $\sim 200-300$ ms after bounce, the strong explosion rapidly inhibits further accretion onto the PNS, beyond which the GW signal is somewhat broadband and lasts until $\sim 800$ ms. Most of the emission occurs near frequencies of $\sim 300-500$ Hz.
\end{itemize}

\subsection{General procedure \label{subsec:gen_proc}}

\begin{sidewaysfigure*}
    \centering
    \vspace*{9cm}\includegraphics[width=\textheight]{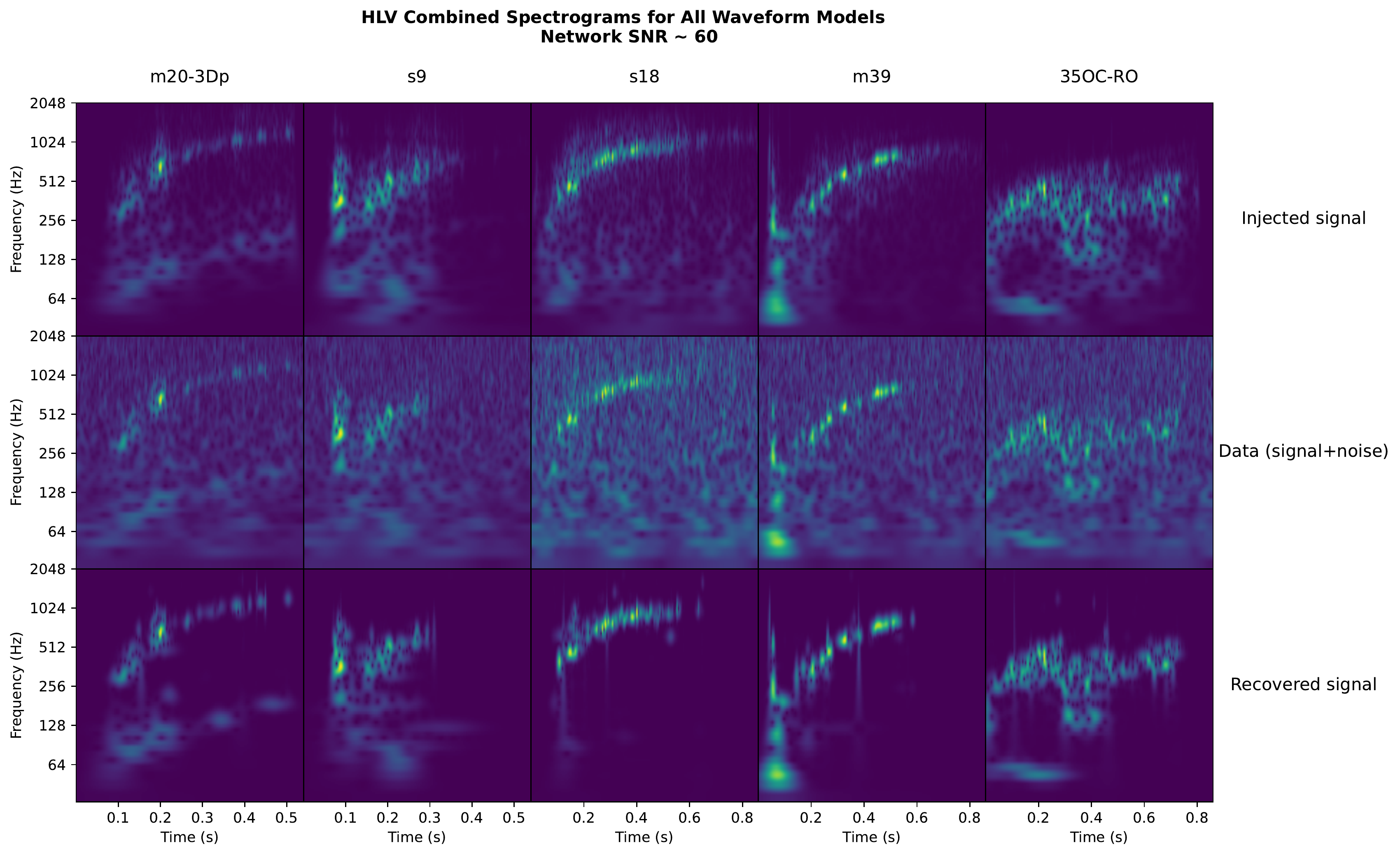}
    \caption{The combined Q-transform spectrograms for the three detectors for the injected signal waveform (top row), the data that BayesWave analyzes (injected signal in simulated gaussian detector noise, middle row), and the BayesWave recovered waveform (bottom row). Out of the 100 injections for each waveform model, one representative case with a relatively high network SNR $\sim 60$ is shown. We can see that for such loud signals, in each case BayesWave manages to recover all dominant features visible in the spectrogram, with overlap accuracies of $\sim 80\%$ for all waveforms. This includes the high-frequency evolving g or f mode track from the PNS oscillation, which is most easily identifiable in the s18 and m39 models, as well as the bounce and prompt convection signal which is visible in the s9 and m39 models. The 35OC-RO waveform is the most distinct with a relatively broadband track since the explosion is magneto-rotationally driven instead of due to neutrino heating. Note that the time-scale for the m20-3Dp and s9 waveforms is shorter than the rest ($\sim 0.5$ s compared to $\sim 0.8$ s).}
    \label{fig:spectrograms}
\end{sidewaysfigure*}

For each waveform model we first take the GW signal as generated by the corresponding simulations in the source frame and project them into the detector frame based on a given set of extrinsic parameters. This set of parameters includes the sky position of the supernova, i.e. its right ascension and declination, the distance to the source, the orientation of the source relative to the line of sight, the polarization angle of the signal, and the time of observation, which influences the detector antenna response pattern in relation to the sky position.

We generate two population sets for these parameters for each model: in the first set the sky positions follow a random uniform distribution across the whole sphere, the source orientations and polarization angles are uniform randomly distributed across all physically possible values, the time of observation is set to be 10 seconds apart (for a uniform-in-sky population the observation time makes no practical difference, and we choose this time for convenience), and the distances are then determined such that the resulting signal in the detector frame is log-uniform randomly distributed in SNR from 10 to 100. We produce 100 distinct signals to analyze in this way. In the second population set we change the sky positions and distances to model after a realistic spatial distribution of stars in the Milky Way galaxy, as done in \citet{LVK2021arXiv}. The Galaxy is modeled as a combination of the bulge, thin disk and thick disk with parameters from the best-fitting contracted NFW halo model in \citet{Cautun2020}. The time interval between signals is increased to 10 minutes, and a total of 150 detector signals are thus generated spanning approximately 24 hours. This is done since the detector sky antenna sensitivity pattern changes on the time scale of a day due to the rotation of the Earth, and we must account for the fact that the sources are not uniformly distributed on the sky, but mostly lie within a narrow band that traces the Galactic disk.

These selected signals are then injected (added) into simulated Gaussian noise that is generated based on the expected design spectral noise sensitivity of the Advanced LIGO \citep{aLIGO2015} and Advanced Virgo \citep{AdVirgo2015} detectors, taking into account the relative positions and orientations of the individual Hanford, Livingston, and Virgo detectors. The end result is a GW data file for each injected signal that has three separate channels for the strain data in each detector, mirroring what we would expect to analyze for a real signal during an observing run.

Finally, we analyze each of these signal-plus-noise data files using the BayesWave algorithm, performing waveform reconstruction of the supernova GW signal within the simulated detector noise (the injection-reconstruction process is illustrated in Fig. \ref{fig:spectrograms}). Since BayesWave is based on a Bayesian approach to model the signal waveform and it samples the posterior distribution of the waveform parameters, for each run the output is a distribution of recovered waveforms, where each waveform represents one step in the sampler. For the runs in our study BayesWave performs a total of 4 million RJMCMC sampler iterations. To keep computational costs reasonable we then sub-select and save every hundredth sample in the latter half of the run (i.e. once the distribution is steady). The BayesWave output products are then calculated based on these 20000 waveforms, and the median recovered waveform (as described in \citet{Cornish2021}) is used as the representative final waveform. Similarly, for the network overlap we quote the median network overlap value of the 20000 waveforms, and show the $1\sigma$ standard deviation of that distribution as a measure of uncertainty on the final value wherever possible.

\subsection{Optimizations}

We aim to optimize the BayesWave run settings to maximize waveform recovery from supernovae for the different waveform models considered and across the range of possible extrinsic signal parameters. We characterize differences in performance between different run settings using the network overlap parameter, where an increase in the overlap indicates a positive gain. We consider this to be a more robust measure of whether the changes we introduce have a positive or negative impact, as compared to measuring the SNR of the recovered waveform. This is because the recovered SNR increases as BayesWave places more wavelets, regardless of whether those wavelets are capturing features appropriately. The overlap, however, appropriately penalizes any frivolously placed wavelets, focusing not on simply the excess power in the recovered signal, but how accurately that excess power matches the shape of the true injected signal.

Since in the case of a real detection we will not know \textit{a priori} what kind of explosion model is the underlying cause of the signal, any optimization efforts we make must be robust across all five different waveforms considered, and for the entire SNR range studied. From the potential optimizations we consider, we find two significant improvements based on: i) knowing accurate sky localization information from electromagnetic and neutrino counterparts (e.g. \cite{Nakamura2016,AlKharusi2021}), and ii) increasing the allowed wavelet quality factor $Q$, which is the number of cycles of the wavelet over one e-folding of the Gaussian envelope (i.e. describes how spread out in time it is). Thus our optimized runs correspond to setting a fixed sky location and $Q_{max} = 100$, compared to the initial un-optimized runs with default settings in which the sky location is not fixed (uniform prior over the whole sky) and $Q_{max}=40$.

We also explored using frequency-evolving `chirplets' \citep{Millhouse2018} instead of the fixed-frequency wavelets. However, this approach did not yield consistently positive results, and we discuss this in more detail in Section \ref{sec:discuss}. We also note that a recent update to the BayesWave algorithm enables relaxing polarization constraints from elliptical polarization to generic polarization \citep{Cornish2021}. We were unable to fully explore this new feature for our present work, and leave a more robust analysis of its potential for CCSN signals to future studies.

Along with this we find that in order for the comparison to be appropriate and for the BayesWave runs to converge in all cases, we have to be careful of a few other run parameters. The analyzed segment length should not be less than 4 seconds, as below that BayesLine \citep{Littenberg2015} can not determine the spectral density of the noise accurately. The analysis window where BayesWave is allowed to place signal wavelets should be kept to approximately 1 second to make sure no part of the potential signal is being missed, and be centered on the duration of the signal (in a real scenario the `trigger' time of the event can be estimated either from low-latency GW search algorithms like coherent WaveBurst \cite{Klimenko2016}) or from electromagnetic or neutrino counterpart detections of CCSNe. The maximum number of allowed wavelets should scale with the SNR of the signal, where we had to set this limit to 150 wavelets for the highest SNR signals in this study. The prior on the number of wavelets should also be specified to be uniform across the allowed window of \{1, 150\}. The number of iterations in the RJMCMC must also not be less than 4 million, as the sampler can sometimes take at least this many steps to converge to a steady distribution, especially for higher SNR CCSN signals. Finally, we also recommend keeping the allowed frequency range wide, between $\sim 16-32$ Hz at the lower end to $\sim 2$ kHz (i.e. a maximum sampling rate of 4 kHz) at the higher end, to account for most of the features we can expect to observe from CCSN GW waveform models with current detectors.

\section{Results} \label{sec:results}

In all subsequent sections we present our main results and figures based on the optimized BayesWave run settings (fixed sky location, $Q_{max} = 100$). We first show how accurate the BayesWave reconstructions are for the different types of CCSN waveform models and how this varies based on the source parameters. We then quantify and describe the increase in overlap accuracy we are able to achieve when using the optimized runs compared to the initial un-optimized runs (uniform sky prior, $Q_{max} = 100$). This is followed by the BayesWave CCSN signal reconstruction prospects presented in the context of a population of sources modeled after the Milky Way stellar distribution.

\subsection{Accuracy of BayesWave reconstructions}

To show the potential capability of BayesWave in recovering a CCSN waveform and all its features if the signal is loud, one instance of the injection-reconstruction process and the median recovered waveform for each model is shown in Fig. \ref{fig:spectrograms}. For each model we consider all the injections in the SNR range 55-65, and then select the one which has the median overlap value within that sample as a representative example. The features are most clearly distinguished in time-frequency spectrograms, which are plotted in Fig. \ref{fig:spectrograms} using a constant-Q transform \cite{Brown1991} ($Q=16$). Since our reconstructions are based on the three detector HLV network and BayesWave simultaneously computes the recovered waveform in all three detectors, we sum the power in the individual spectrograms after accounting for the appropriate time delay shift, and show the HLV combined spectrograms (noting that this coherent sum emphasizes the signal more over the background noise when compared to single detector signal+noise spectrograms). At such high SNRs, we can see that BayesWave is recovering all previously identified GW emission features, including the rising g or f mode PNS oscillation track visible in all models, the early broadband prompt convection in e.g. m39, and the low-frequency SASI in m20-3Dp.

\begin{figure*}
    \centering
    \includegraphics[width=\textwidth]{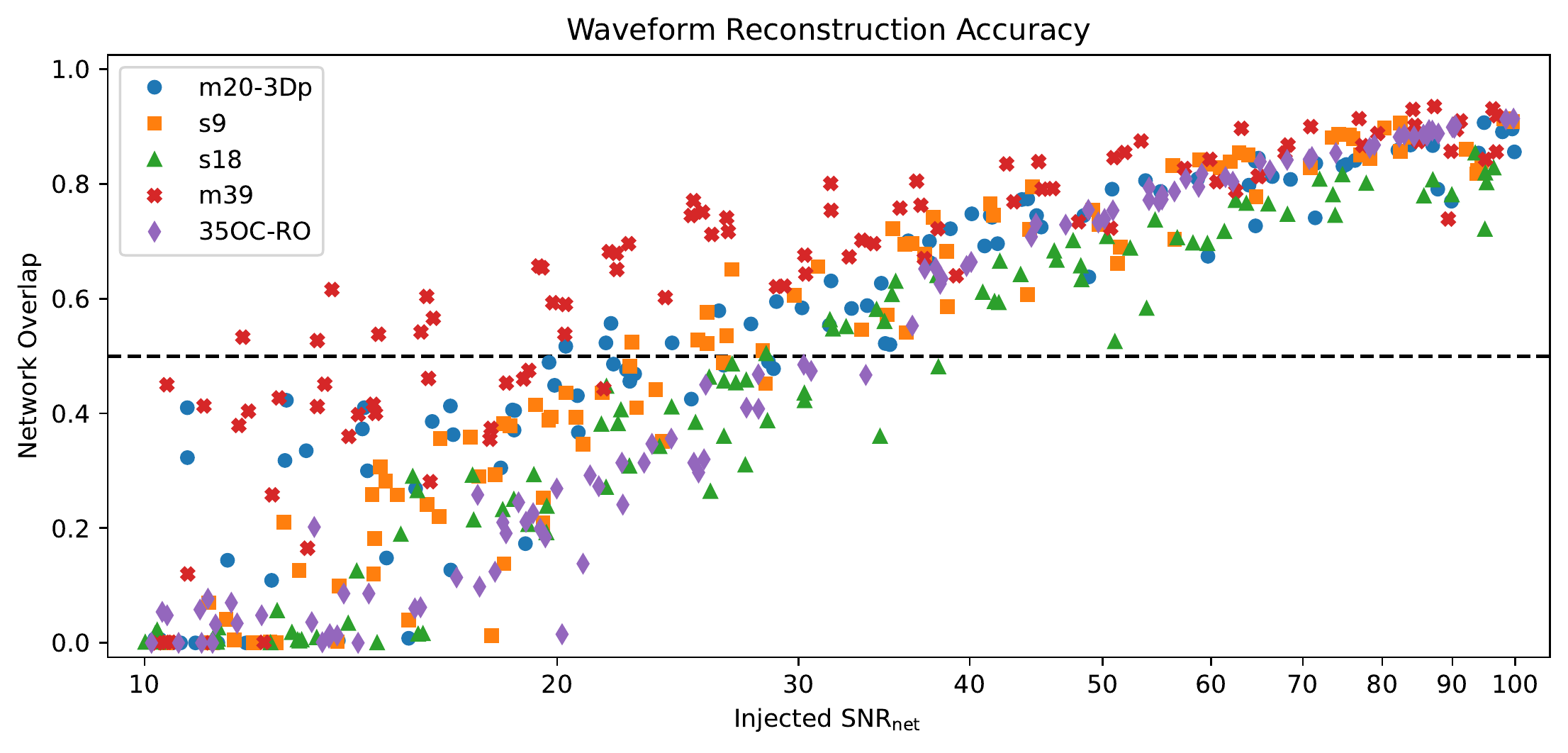}
    \caption{A scatter plot of the BayesWave reconstructions for all the injections in the study, showing the median accuracy of the recovered waveform as measured by the network overlap (match) parameter plotted against the network SNR of the injected signal. Uncertainty error bars are omitted here for readability (see Fig. \ref{fig:overlap_vs_distance}). The dashed black line indicates the 50\% accuracy threshold. The reconstruction accuracy between models varies greatly at low SNRs, with a network SNR of $\gtrsim 30$ required to exceed the 50\% threshold for all models. The accuracy converges to $\sim 90\%$ for the highest SNR signals in this study. The differences between models can be explained by the differences in the waveform features and how compact they are in time-frequency. The scatter in the overlap accuracy for a given model near a specific SNR is a result of the small changes in the waveform morphology given varying extrinsic parameters (such as inclination angle).}
    \label{fig:overlap_vs_snr}
\end{figure*}

Fig. \ref{fig:overlap_vs_snr} shows the overlap accuracy of the BayesWave reconstructed waveform for all the supernova GW signal injections analyzed in our study. The median network overlap (match) for each reconstruction in the optimized runs is plotted against the injected signal SNR for the 3 detector HLV network (root-sum-square of the individual detector SNRs). As observed in previous studies (e.g. \cite{Szczepanczyk2021}), the accuracy of the reconstruction generally increases as the signal SNR increases. Table \ref{table:50_percent_overlap} lists the network SNR required for each model for the network overlap to exceed 0.5 and the majority of the waveform be recovered by BayesWave. Overall, when the SNR $\gtrsim 30$, all explosion models are confidently recovered.

\begin{table}
\renewcommand{\arraystretch}{1.2}
\centering
\begin{tabular}{ c c c }
 \hline
 \hline
 \textbf{~Waveform Model~} & $\mathbf{~SNR_{net}~}$ & \textbf{~Distance (kpc)~} \\
 \hline
 \hline
 m20-3Dp & 25 & 1.2 \\
 \hline
 s9 & 25 & 0.6 \\ 
 \hline
 s18 & 30 & 5 \\ 
 \hline
 m39 & 20 & 40 \\ 
 \hline
 35OC-RO & 30 & 45 \\ 
 \hline
 \hline
\end{tabular}
\caption{The approximate GW signal network SNR value or source distance required for the BayesWave reconstruction network overlap value to be $O_{net} \approx 0.5$ for each model. For signals with higher SNRs or lower source distances than those listed the reconstruction will have $O_{net} > 0.5$, i.e. the majority of the waveform will be accurately recovered by BayesWave.}
\label{table:50_percent_overlap}
\renewcommand{\arraystretch}{1}
\end{table}

However, similar to what prior work by \citet{Gossan2016} and \citet{Szczepanczyk2021} found for other analysis pipelines, there is a difference in BayesWave's reconstruction performance between the different explosion models, especially at lower SNRs. We can see that the s18 and 35OC-RO models have the lowest accuracy, the m20-3Dp and s9 models somewhat higher, and the m39 model has the highest reconstruction accuracy (these trends are also true for the un-optimized run results which aren't shown). At higher SNRs these differences become less pronounced, with reconstruction accuracies reaching $\sim 90\%$ at SNRs $\sim 100$. The differences can be explained by the different waveform morphologies. How the power is distributed in time-frequency governs how well the features can be captured by BayesWave's Morlet-Gabor wavelet basis, with well-localized signals expected to be better reconstructed.

For each explosion model we also see that the points near a given SNR for that model have a small scatter due to different extrinsic parameters. Waveform features that BayesWave can more accurately reconstruct are a little bit more pronounced for some combination of extrinsic parameters than others that still give a similar SNR. For example, for the m39 model we know that the prompt convection signal right after core bounce is most readily observed when the line of sight is aligned with the star's equatorial plane (inclination angle $= 90^{\circ}$) as opposed to the pole \citep{Powell2020}. And since this feature is more compact in time-frequency as compared to the f-mode PNS oscillation track, BayesWave can recover the equatorial signals better than the polar ones.

\begin{figure*}
    \centering
    \includegraphics[width=\textwidth]{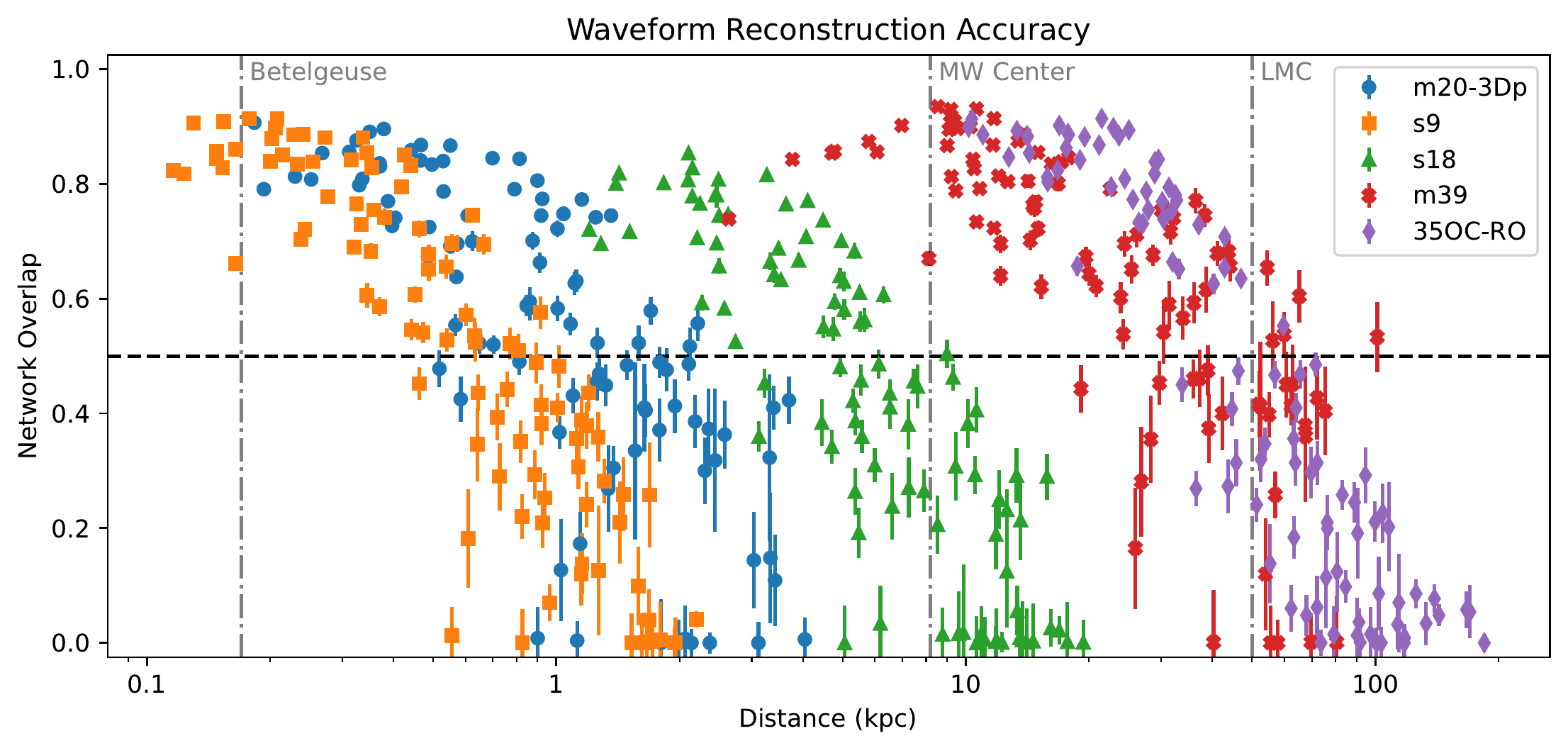}
    \caption{The accuracy of the BayesWave recovered waveform as a function of the source distance (and explosion model) with $1\sigma$ uncertainty error bars. The dashed black line indicates the 50\% accuracy threshold. Also shown for reference are three benchmark distances: Betelgeuse, the Milky Way center, and the center of the Large Magellanic Cloud. The distances at which the waveform is accurately recovered by BayesWave and passes the 50\% threshold varies by almost two orders of magnitude between the different waveform models. For the low mass and relatively low explosion energy models of m20-3Dp and s9 this occurs at $\sim 1 ~\mathrm{kpc}$, for the typical neutrino-driven explosion model of s18 this is closer to the Milky Way center, and for the high explosion energy models with rapidly rotating progenitors the recoverable distances are up to the LMC at $\sim 50 ~\mathrm{kpc}$.}
    \label{fig:overlap_vs_distance}
\end{figure*}

Following the approach in \citet{LVK2021arXiv}, to find out how far away a supernova source can be for BayesWave to be able to detect and reconstruct the signal in Advanced LIGO-Virgo, we show the reconstructed overlap accuracy for the injections analyzed plotted against the source distance in Fig. \ref{fig:overlap_vs_distance}. Since each waveform model is based on very different progenitor stars and uses different modeling techniques, the total amount of energy released in the explosion, and thus in GWs, varies over a few orders of magnitude. The source distance required to have a high amplitude GW signal reach the LIGO-Virgo detectors is then significantly dependent on the waveform model considered (as was previously found \cite{Gossan2016, Szczepanczyk2021, LVK2021arXiv}). Table \ref{table:50_percent_overlap} also summarizes the source distances at which the BayesWave reconstructed signal has an overlap accuracy of $\sim 0.5$ for each waveform model, showing an almost two orders of magnitude difference between the low mass neutrino-driven explosion of the s9 model and the magnetorotationally driven explosion of the 35OC-RO model.

\subsection{Optimization gains}

\begin{figure*}
    \centering
    \includegraphics[width=\textwidth]{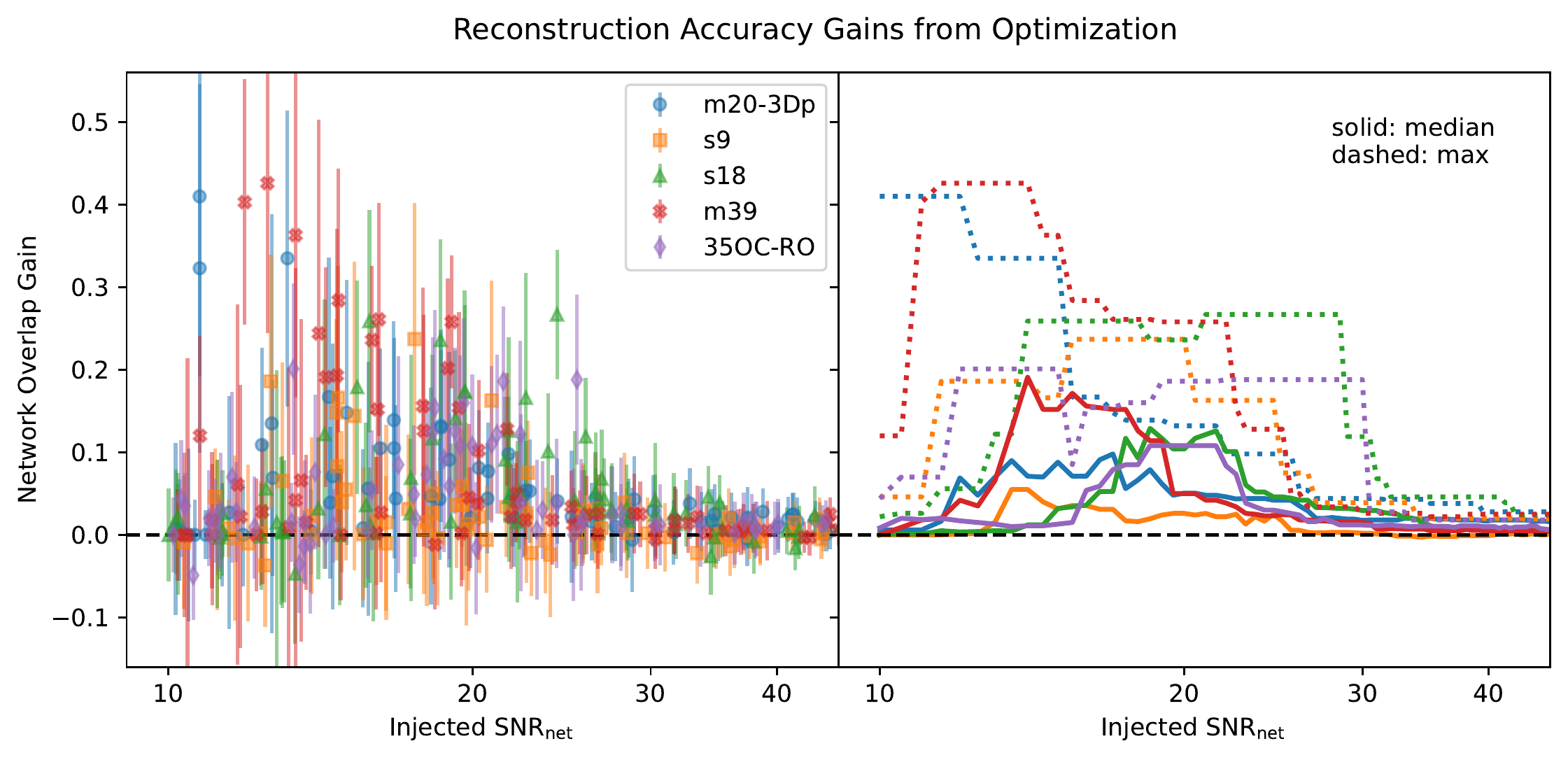}
    \caption{The gain in the network overlap of the recovered waveform when using the optimized BayesWave run settings as compared to the initial un-optimized runs, plotted against the injected signal SNR. The scatter plot on the left shows the gains for each individual injection-reconstruction, along with the $1\sigma$ uncertainty error bars on the gain. The gains are either positive or consistent with zero. The plot on the right summarizes this information showing the median overlap gain (solid lines) and the maximum gains observed (dashed lines) near a given SNR. The median gains are on the order of $\sim 10\%$ in the low SNR regime of $\sim 15-20$, and consistent with zero for SNR $\gtrsim 30$. The maximum observed gains can be quite large and significant, reaching an increase in overlap accuracy of about 40\% for the m20-3Dp and m39 models, and more than 20\% for the other three models. }
    \label{fig:overlap_optimization_gains}
\end{figure*}

In order to quantify the effects of our BayesWave optimization efforts, and to see in which cases they make the biggest impact, in Fig. \ref{fig:overlap_optimization_gains} we plot the gain in reconstruction accuracy of the waveform between the initial and optimized BayesWave runs against the network SNR of the injected signal. The gains are most pronounced in the low SNR regime of $\sim 15-20$, where they are typically on the order of $\sim 10\%$. However, the maximum gains observed across our study can be very high, from $\sim 20\%$ for the 35OC-RO model to $\sim 40\%$ for the m20-3Dp and m39 models. This SNR regime corresponds to the transition between cases where the signal is too low in amplitude for BayesWave to distinguish much of it from the detector noise, and cases where the signal is loud enough that BayesWave is starting to confidently recover most of the waveform. Thus the optimizations are most impactful for the marginal cases, and have no significant impact for signals with $\mathrm{SNR} \gtrsim 30$.

\begin{sidewaysfigure*}
    \centering
    \vspace*{9cm}\includegraphics[width=\textheight]{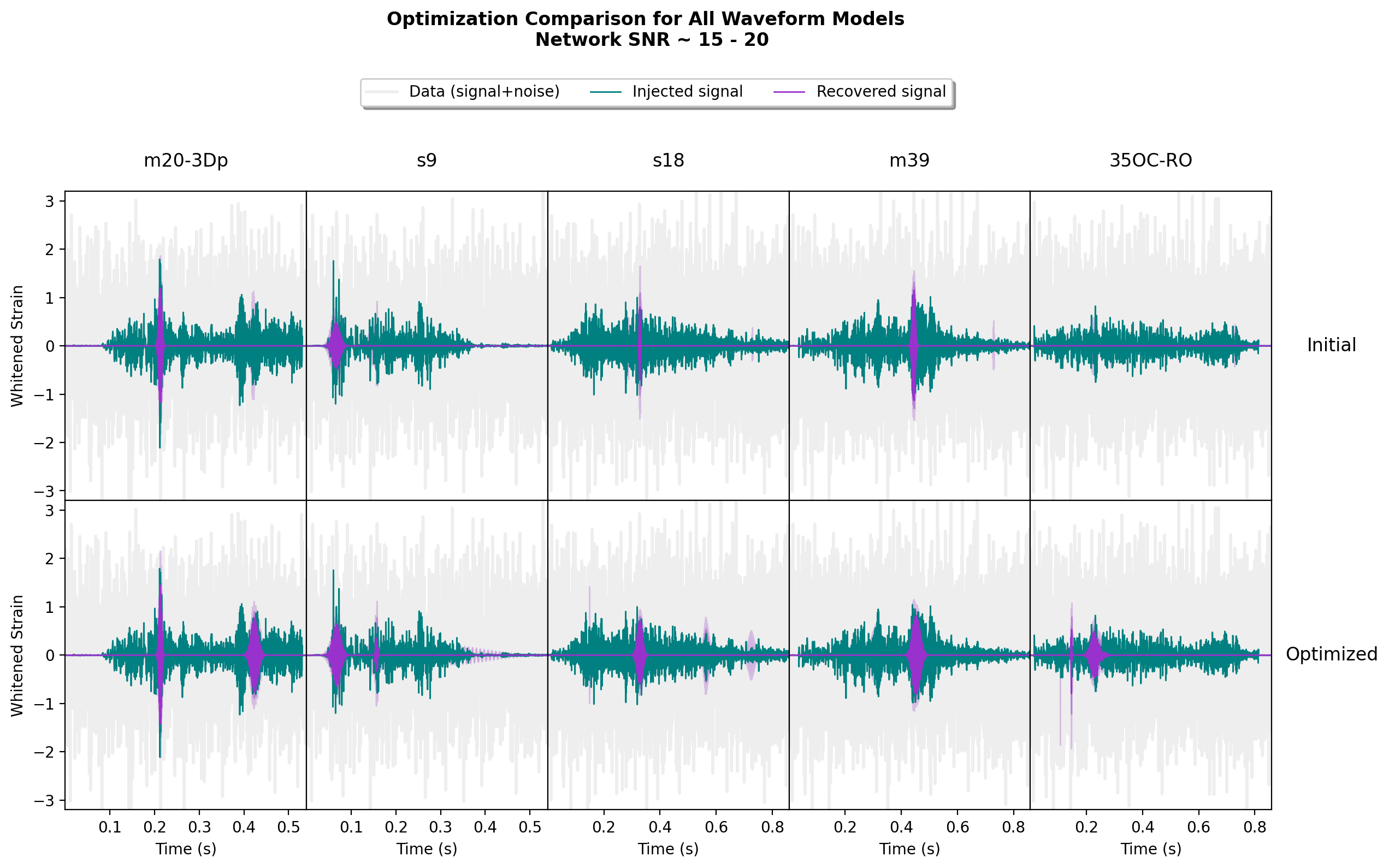}
    \caption{The supernova waveform signals and their reconstructions in the time domain, showing the comparison between the initial BayesWave reconstructions and the optimized version. Plotted in darkest purple is the median reconstructed waveform, while the lighter shaded bands are the 50\% and 90\% credible intervals. These are overlaid on top of the true injected signal, plotted in teal. To get a sense of how this signal would look like in one of the LIGO detectors, the combined signal plus detector noise is shown in gray in the background. Out of the 100 injections analyzed for each waveform model, one representative case in the low network SNR regime ($\sim 15-20$) is shown. This is the SNR range where the optimization leads to the highest gains in reconstruction accuracy, as seen in Fig. \ref{fig:overlap_optimization_gains}. We can see that the gains in reconstruction can generally be attributed to BayesWave's ability to more confidently and accurately capture the high-amplitude features. Note that the time-scale for the m20-3Dp and s9 waveforms is shorter than the rest ($\sim 0.5$ s compared to $\sim 0.8$ s).}
    \label{fig:waveforms_optimized}
\end{sidewaysfigure*}

A more qualitative picture of how these optimizations manifest is shown in a comparison of the time series reconstructions of the waveforms in Fig. \ref{fig:waveforms_optimized}. For signals with $\mathrm{SNR} \sim 15-20$ we choose the case which has the median gain in overlap within that SNR range as a representative example for each model, and show the injected and BayesWave reconstructed signals in one of the LIGO detectors (Hanford and Livingston have the same expected design sensitivity \citep{aLIGO2015}, so they are qualitatively interchangeable in our study). We can see that the gains in reconstruction for these examples can be attributed to a combination of BayesWave being able to: i) find the highest amplitude part of the signal when it wasn't able to recover anything before (35OC-RO), ii) recover the same part of the signal it was before but more accurately (m39 and s9), and iii) recover additional high-amplitude parts of the signal compared to only the highest amplitude portion before (s18 and m20-3Dp).

\subsection{Prospects for a Milky Way source distribution}

\begin{figure}
    \centering
    \includegraphics[width=\columnwidth]{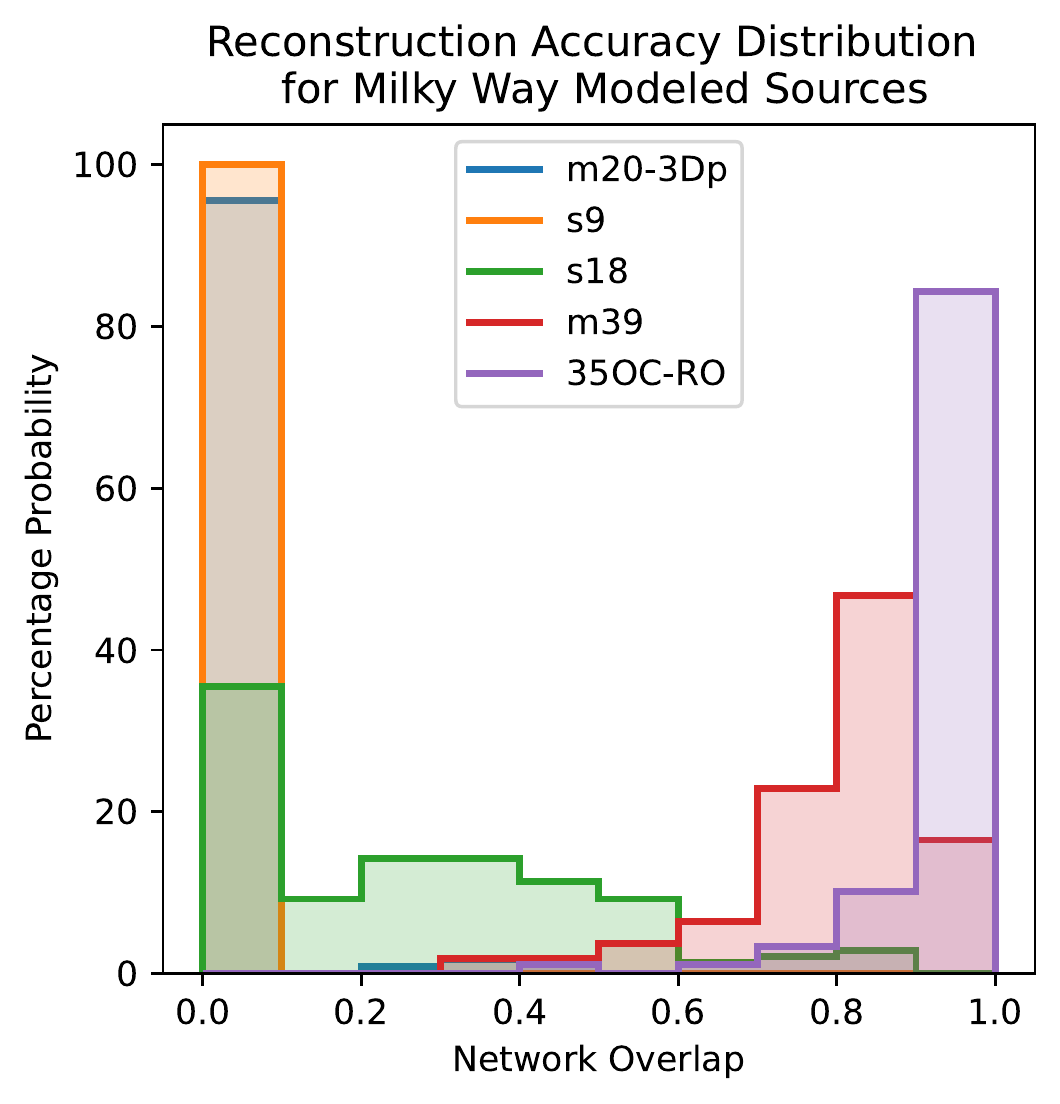}
    \caption{The histogram distribution of the optimized BayesWave reconstructed network overlap for each supernova model, for a population of injections with extrinsic parameters modeled according to a realistic distribution of stars in the Milky Way. We see that for such a distribution, we do not expect to be able to recover the GW waveform with BayesWave from low energy m20-3Dp- or s9-like supernova models. For more typical neutrino-driven explosion models like s18, BayesWave can recover a non-zero overlap in many cases, but high accuracy reconstructions with overlap $> 0.5$ occur only in $\sim 15\%$ of cases. For stronger explosions in the Galaxy from rapidly rotating progenitors (m39, 35OC-RO), almost all of them can be expected to be confidently and accurately recovered.}
    \label{fig:galactic_overlap_distribution}
\end{figure}

We show the prospects for accurate recovery of each model waveform using a realistic distribution of CCSN sources within the Milky Way galaxy, modeled according to its stellar distribution, in Fig. \ref{fig:galactic_overlap_distribution}. This figure shows the histogram distribution of the network overlap in the optimized BayesWave runs for Milky Way distributed signals.

We see that for the low explosion energy models of m20-3Dp and s9 we can not expect to reconstruct the GW signal in almost all cases (the overlap is non-zero for $<5\%$ of sources), consistent with what \citet{LVK2021arXiv} find. This is primarily because the stellar distribution in the Galaxy is most concentrated towards the central bulge near 8 kpc, while we see in Fig. \ref{fig:overlap_vs_distance} that these models are not confidently recovered beyond $\sim 1-2$ kpc.

For the s18 model we have a somewhat broader distribution, with a non-zero overlap in $\sim 65\%$ of cases, and a confident reconstruction with overlap $> 0.5$ for $\sim 15\%$ of the sources. The most likely (non-zero) overlap value, coincident with sources at the Galactic center, is $\sim 0.3$. Due to the fact that our optimizations were largely effective at lower SNRs, and this SNR regime is coincident with near the Galactic center for the s18 model, it is also the only model for which we see a measurable increase in the percentage of cases that are recovered between the un-optimized and optimized runs: a 5\% increase from 11\% to 16\% for confident reconstructions, and a 15\% increase from 52\% to 67\% for non-zero overlap.

For the two high explosion energy models we see that $> 95\%$ of cases have overlap $> 0.5$. The most likely overlap, coincident with sources at the center of the Milky Way, is $\sim 0.85$ for m39 and $\sim 0.95$ for 35OC-RO. In Fig. \ref{fig:galactic_spectrograms} we also show spectrograms for the most likely reconstruction for sources at the Galactic center for each waveform model.

\section{Discussion} \label{sec:discuss}

\subsection{Reconstruction accuracy model dependence}

The duration of the GW emission plays a central role in influencing the quality of the BayesWave reconstruction. Since the SNR is an integrated measure of the signal power across the analyzed segment, when comparing different model injections at a fixed SNR, their total power is the same. However, the injected signals in Fig. \ref{fig:spectrograms} show that significant GW emissions last for about 0.5 seconds for s18 and about 0.7 seconds for the 35OC-RO model, making them more spread out in time compared to the $\sim 0.3$ seconds emissions from m20-3Dp and s9. On average, this effectively distributes the signal power so that analyzing a similar 0.3 second segment in the longer models would have about half the power as compared to the shorter ones. Since BayesWave places wavelets in time-frequency trying to capture this localized power wherever it confidently can \citep{Cornish2015}, the wavelet basis better captures signals that have their excess power concentrated in a shorter time (following the same reasoning why high mass compact binary coalescences are better reconstructed compared to low mass ones in \citet{Ghonge2020}). For the m39 model, even though it is one of the longer duration models with significant GW emissions lasting about 0.6 seconds, depending on the source orientation it often produces much of that excess power in the short duration early rotational bounce signal (seen in Fig. \ref{fig:spectrograms}), which BayesWave easily recovers.

Compactness in frequency of the GW emission further explains differences in reconstruction accuracy. For example, in the m39 model the f-mode PNS oscillation track is very narrow in time-frequency. Compare this to the s18 model where no high amplitude prompt convection mechanism contributes and the signal energy is distributed throughout a relatively broader PNS g-mode oscillation track (as well as what seems to be a low-frequency persistent background component), making it harder for BayesWave to recover. The other three models are similarly more spread out in frequency than m39, negatively affecting the ability of the compact Morlet-Gabor wavelets to capture this more broadband power.

At lower SNRs these differences in the reconstruction accuracy are more pronounced because BayesWave needs the localized SNR in each segment of the signal to be above a certain threshold to justify placing a wavelet there (since it takes a Bayesian approach, the increase in signal likelihood must exceed the Occam factor penalty for increasing the signal model complexity \citep{Cornish2015,Kanner2016,Lee2021}). Capturing localized excess is then only possible close to where the signal amplitude peaks for lower SNR signals (as seen in Fig. \ref{fig:waveforms_optimized}), which varies significantly from model to model. On the other hand, for high SNR signals the signal amplitude is high enough to be distinguished from the noise across the whole waveform, and so BayesWave can place wavelets across the signal relatively unaffected by the specific distribution of the power.

In future work we plan to do a deeper investigation on the possibility of distinguishing between different model features and extracting physical parameters of the progenitor using machine learning techniques, based on the waveform reconstructions that BayesWave outputs.

\subsection{Reconstruction distances in context}

In Fig. \ref{fig:overlap_vs_distance} along with presenting the distance dependence of the BayesWave reconstructions for the optimized runs, we also include distances to some common benchmarks for comparison, similar to Figure 7 in \citet{LVK2021arXiv}. Betelgeuse is the nearest red supergiant to Earth, expected to end its life in a CCSN \citep{Dolan2016}. The Milky Way center is the region with the highest density of massive stars in the Galaxy, and so is the most likely source for the next Galactic supernova. The Large Magellanic Cloud is the host galaxy to supernova 1987A, one of the most recent and closest CCSN observed, and the only one also detected with neutrinos \citep{Bionta1987, Hirata1987, Alexeyev1988}. We see that for the low explosion energy models of s9 and m20-3Dp, the GW waveform is very well recovered by BayesWave at distances close to Betelgeuse, but the accuracy falls below the 50\% threshold very quickly beyond $\sim 1-2$ kpc. The typical neutrino driven explosion for s18 is recovered well at galactic distances, up to the center of the Milky Way, but can not be recovered for sources beyond our galaxy. For the very highest explosion energy models of m39 and 35OC-RO however, the BayesWave reconstruction has accuracies of $\sim 90\%$ close to the Milky Way center, and is well recovered even up to the LMC. If Betelgeuse goes supernova during the advanced detector era then we can see that BayesWave will accurately recover and reconstruct almost all of the signal, regardless of waveform model (even if the explosion is not successful, as in the m20-3Dp model).

These distances are also significantly larger than the detection distances found for the same set of waveform models studied with BayesWave in \citet{LVK2021arXiv}, as shown in Figure 7 of their paper. This is expected as the LVK study is based on the actual detector noise curves recorded for the LIGO and Virgo detectors during the third observing run, which is roughly a factor of 1.5 less sensitive than the design Advanced LIGO-Virgo configuration that we use in this study, in anticipation of the fourth observing run \citep{LVK2020}. Note that there is also a difference in the distance measure analyzed: the LVK study quotes the distance at which 50\% (and 10\%) of the total injections are detected, whereas in our study we look at the distance at which the recovered waveform accuracy is on average 50\%. We also find that the BayesWave reconstruction overlap accuracies at given distances (and SNRs) for the advanced LIGO-Virgo detectors are broadly consistent with those reported by \citet{Szczepanczyk2021} using the coherent WaveBurst algorithm.

As was done for the third observing run sensitivity in \citet{LVK2021arXiv}, for a realistic distribution of CCSN progenitors within the Milky Way we have also quantified the prospects for BayesWave to confidently reconstruct the emitted GW signal for the expected sensitivities in the fourth observing run (Fig. \ref{fig:galactic_overlap_distribution}). This also demonstrates how the distribution of sources within the Galaxy modulates the detectability of each supernova model, as most stars are concentrated near the center of the Galaxy $\sim 8$ kpc away from the Sun. Should the next Galactic CCSN be a low energy or failed explosion, we do not anticipate being able to detect it in Advanced LIGO-Virgo with BayesWave (unless we are very lucky and the source is $\lesssim 1$ kpc away). If it is a typical neutrino driven explosion like the s18 model however, we expect that there is a $\sim 65\%$ chance BayesWave will recover a part of the GW signal, and a $\sim 15\%$ chance it will be confidently reconstructed. Finally, if the explosion follows from a rapidly rotating progenitor, then BayesWave will almost certainly recover the GW features confidently ($> 95\%$ probability). However, such rapid rotation is only expected to occur in $\sim 1\%$ of all CCSN progenitors \citep{Woosley2006}.

\subsection{Limits to BayesWave reconstructions}

The work by \citet{Pannarale2019} characterizes BayesWave's performance in general from a theoretical approach, and suggests a formulation for the maximum possible overlap (match) in Equation 8 of their paper. We test this prediction against our CCSN reconstructions for all five of the waveform models studied, and plot the difference between the two for our optimized runs in Fig.~\ref{fig:overlap_prediction_comparison} (included in the appendix). What we see is similar to what \citet{Pannarale2019} find, which is that more complicated and longer waveforms (such as the 35OC-RO in our study) show a larger disparity with the predictions and have a harder time reaching that theoretical maximum as compared to waveforms that are more compact in time-frequency (such as the m39 waveform in our study). This also explains why BayesWave has a harder time recovering the signal accurately from all of the supernova models in our study when compared to it's performance in recovering the signal from coalescing binary black holes as studied in \citet{Pannarale2019}.

The disparity between the predicted overlaps and the CCSN overlaps in our study also show an SNR dependence, where the predictions overestimate much more at lower SNRs than at higher. This happens because \citet{Pannarale2019} state Equation 8 is valid in the limit that the squared SNR recovered per pixel (wavelet) is much greater than unity, which is a requirement that holds true for reconstruction of high SNR signals in our study but becomes more tenuous for lower SNR signal reconstructions. Theoretically we are expecting that there should be some useful information, but practically at such SNRs the signal is buried in Gaussian noise and BayesWave has a very hard time extracting any part of it, giving very little or no information.

We also consider the limits of BayesWave to understand why our run optimization efforts work well at relatively lower SNRs. When the signals are not very loud, BayesWave's certainty in determining the time delay of certain signal features between the three different detectors decreases, and so the reconstructed sky localization is not accurate, with the 90\% confidence sky area not constrained to better than $\sim 50 ~\mathrm{deg^2}$ for SNR $\lesssim 30$. This increase in the posterior distribution for the source sky position parameter is associated with a decrease in the calculated likelihood when placing a wavelet to capture that signal feature, and so in the RJMCMC BayesWave places a wavelet to capture that power less often, resulting in a lower median overlap. When the true sky position is known and fixed however, the parameter space is significantly reduced, and so BayesWave places wavelets at the corresponding signal feature locations much more confidently and thus much more often, increasing the median overlap.

The increase in overlap due to a higher wavelet quality factor is a direct result of the fact that supernova waveforms generally have high amplitude features that are spread out in time, and so longer duration wavelets are required to capture the power not just at the precise time where the amplitude is large, but also in its immediate vicinity (as illustrated in Fig. \ref{fig:waveforms_optimized}). At higher SNRs this makes little difference because instead of having to capture the excess power using only one wavelet, the signal is loud enough that BayesWave can confidently place many wavelets to capture the same power.

In our optimization efforts during our study we had also considered using frequency-evolving `chirplets' \citep{Millhouse2018} instead of the fixed-frequency wavelets, i.e. increasing the allowed waveform complexity in the bid to capture the nuances in the signal morphologies more accurately. However, this was not included in the final optimized BayesWave runs as it did not yield positive gains for all the different waveform models. While using chirplets increased the overlap accuracy by $\sim 5\%$ for low SNR signals for the more broadband 35OC-RO model, it also resulted in a slight net negative overlap at low SNRs for the m20-3Dp and m39 models (while having no measurable impact on the other two models). Combined with the fact that the computing time is almost twice as long for the chirplet model as compared to the wavelet, we recommend that the wavelet model be used for supernova analysis unless the signal is known to be broadband from other analyses.

\section{Conclusion} \label{sec:conclusion}

Core-collapse supernova explosion models predict the production of gravitational waves within the first second after core-collapse, exhibiting dominant emission features with frequencies that lie in the most sensitive regime of current interferometric ground based detectors. Galactic CCSNe represent a rare but promising astrophysical source that could be detected by the Advanced LIGO and Advanced Virgo detectors in one of their upcoming observing runs.

In order to prepare for such a detection, in this paper we study how well the BayesWave algorithm can reconstruct the GW signal emitted by a range of different explosion models with different waveform morphologies, if such a signal is detected by a design sensitivity LIGO-Virgo network configuration. We also optimize the BayesWave run settings for CCSN, focusing on maximizing the reconstruction overlap accuracy, and provide a set of recommendations for the run settings for analysis on a real candidate signal.

We find that BayesWave's reconstruction performance varies depending on the complexity of the waveform, with models that exhibit features that are more compact in time-frequency being better recovered than models that have their power distributed over a longer time and have broadband emission, consistent with previous work \citep{Szczepanczyk2021}. That said, BayesWave can confidently recover the GW signal from a range of supernova explosion models in Advanced LIGO-Virgo for network $\mathrm{SNR \gtrsim 30}$, with maximum accuracies of $\sim 90\%$ for SNR $\sim 100$, as shown in Fig. \ref{fig:overlap_vs_snr}.

The corresponding distances, as seen in Fig. \ref{fig:overlap_vs_distance}, are from as low as $\mathrm{\sim 1 ~kpc}$ for low mass neutrino driven explosions, to up to $\mathrm{\sim 50 ~kpc}$ for high mass magneto-rotationally driven explosions. We put this into context of what percentage of CCSN within our Galaxy we could confidently recover with BayesWave by also studying the reconstructions for a realistic distribution of stellar sources in the Miky Way (Fig. \ref{fig:galactic_overlap_distribution}). We find that should the next Galactic CCSN be a low mass or failed explosion, BayesWave is likely not to detect it. However, there is a $\sim 15\%$ chance the source is close enough that for typical neutrino driven explosions like the s18 model, BayesWave will be able to confidently reconstruct the GW emission. Should a high energy CCSN from a rapidly rotating progenitor be the source, BayesWave will confidently recover the GW signal in $> 95\%$ of cases.

A key result of our work is that optimizing the BayesWave run settings increases the reconstructed accuracy significantly for low SNR signals, with gains of up to $\sim 20-40\%$ for lower SNR signals, and typically $\sim 10\%$ in the SNR $\sim 15-20$ regime, as shown in Fig. \ref{fig:overlap_optimization_gains}. These gains are based on: i) exploiting the multi-messenger nature of CCSN so that the sky location is known from the electromagnetic and neutrino counterpart observations, allowing BayesWave to more confidently recover waveform features, and ii) increasing the allowed wavelet quality factor (to $Q_{max} = 100$) so that high amplitude features that are more spread out in time for supernova signals are more accurately reconstructed. For typical neutrino-driven explosions like the s18 model, this also translates into being able to recover the signal from $\sim 15\%$ more sources in our Galaxy.

Our results show that if there is a CCSN signal detection in Advanced LIGO-Virgo, for loud enough signals the BayesWave algorithm will be able to accurately reconstruct the dominant GW emission features, regardless of the specific model. This will help us distinguish between different supernova explosion models and mechanisms, and allow crucial insight into the dynamics of the very first second of the explosion.

\begin{acknowledgments}
The authors gratefully acknowledge Jade Powell, Meg Millhouse, Katerina Chatziioannou, Tyson Littenberg, Bence B{\'e}csy, Sudarshan Ghonge and Evan Goetz for useful discussions related to this work. The manuscript has benefited from internal LVK collaboration review under the document ID P2100404. This material is based upon work supported by NSF's LIGO Laboratory which is a major facility fully funded by the National Science Foundation. The authors gratefully acknowledge the support of the NSF for provision of computational resources. This research is funded by an NSERC Discovery Grant. GD is supported through the iBOF-project BOF20/IBF/124.
\end{acknowledgments}

\vspace{5mm}

\appendix*

\section{}
\begin{sidewaysfigure*}
    \centering
    \vspace*{9cm}\includegraphics[width=\textheight]{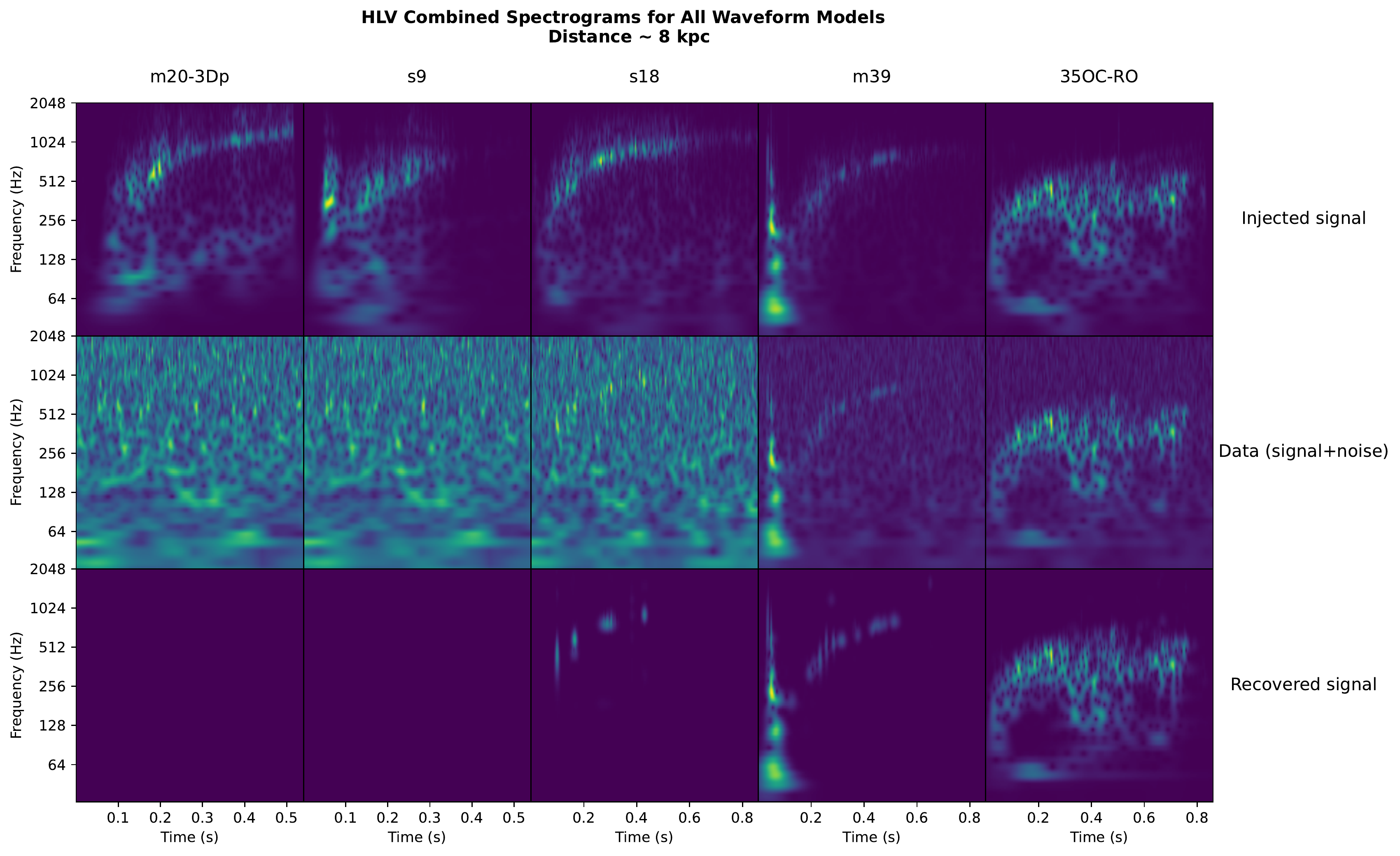}
    \caption{The combined Q-transform spectrograms for the three detectors for the injected signal waveform (top row), the data that BayesWave analyzes (injected signal in simulated gaussian detector noise, middle row), and the BayesWave recovered waveform (bottom row). Out of the 150 Milky Way distributed injections for each waveform model, one representative case at the Galactic center at a distance of $\sim 8$ kpc is shown. The m20-3Dp and s9 waveforms are not reconstructed at all, the s18 model has an overlap reconstruction accuracy of $\sim 30\%$, the m39 model has accuracy of $\sim 85\%$, and the 35OC-RO model has accuracy of $\sim 95\%$. The large differences in accuracy of reconstruction is due to the difference in GW amplitude and energy between models.}
    \label{fig:galactic_spectrograms}
\end{sidewaysfigure*}

To supplement the discussion in the main text, in Fig. \ref{fig:galactic_spectrograms} we show spectrograms for the most likely BayesWave reconstruction for sources near the Galactic center for each waveform model. For each model we consider all the injections from the Milky Way distributed population, in the distance range 7-9 kpc, and then select the one which has the median overlap value within that sample as a representative example.

\begin{figure}
    \centering
    \includegraphics[width=\columnwidth]{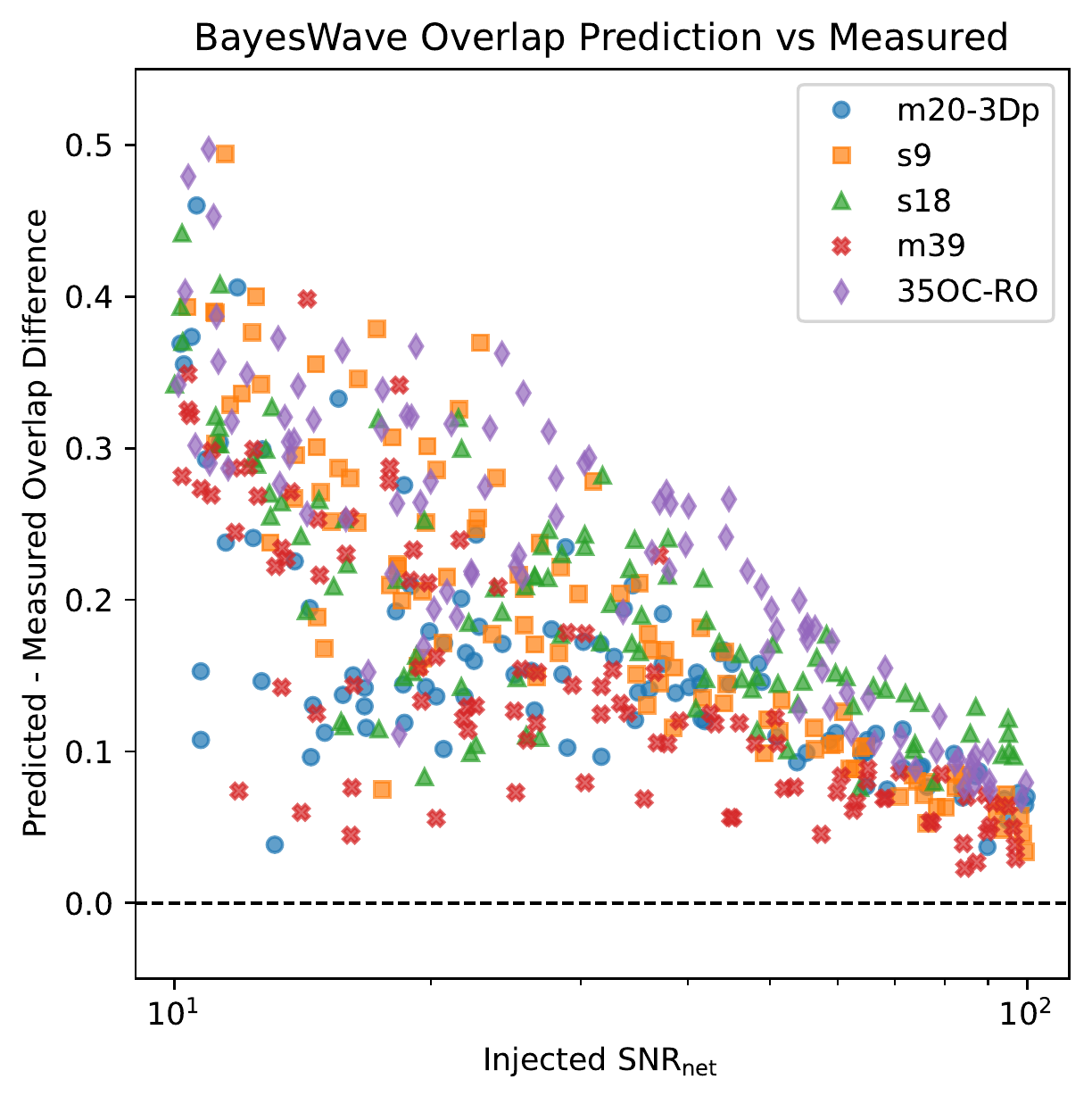}
    \caption{Comparison of the theoretical predictions of the BayesWave overlap according to Equation 8 in \citet{Pannarale2019}, and the measured overlap in our study. Our measurements are worse than the predicted values (as expected), and quantify how the prediction breaks down for longer duration signals like CCSNe. The mismatch is larger for lower SNR signals, and generally larger for models that have more complex features in time-frequency such as the 35OC-RO model.}
    \label{fig:overlap_prediction_comparison}
\end{figure}

We also test the predictions made in the work by \citet{Pannarale2019} characterizing BayesWave's performance from a theoretical approach, against our CCSN reconstructions for all five of the waveform models studied, and the difference between the two for our optimized runs is shown in Fig. \ref{fig:overlap_prediction_comparison}.

\bibliography{ref.bib}

\providecommand{\noopsort}[1]{}\providecommand{\singleletter}[1]{#1}%
\begin{thebibliography}{63}%
\makeatletter
\providecommand \@ifxundefined [1]{%
 \@ifx{#1\undefined}
}%
\providecommand \@ifnum [1]{%
 \ifnum #1\expandafter \@firstoftwo
 \else \expandafter \@secondoftwo
 \fi
}%
\providecommand \@ifx [1]{%
 \ifx #1\expandafter \@firstoftwo
 \else \expandafter \@secondoftwo
 \fi
}%
\providecommand \natexlab [1]{#1}%
\providecommand \enquote  [1]{``#1''}%
\providecommand \bibnamefont  [1]{#1}%
\providecommand \bibfnamefont [1]{#1}%
\providecommand \citenamefont [1]{#1}%
\providecommand \href@noop [0]{\@secondoftwo}%
\providecommand \href [0]{\begingroup \@sanitize@url \@href}%
\providecommand \@href[1]{\@@startlink{#1}\@@href}%
\providecommand \@@href[1]{\endgroup#1\@@endlink}%
\providecommand \@sanitize@url [0]{\catcode `\\12\catcode `\$12\catcode
  `\&12\catcode `\#12\catcode `\^12\catcode `\_12\catcode `\%12\relax}%
\providecommand \@@startlink[1]{}%
\providecommand \@@endlink[0]{}%
\providecommand \url  [0]{\begingroup\@sanitize@url \@url }%
\providecommand \@url [1]{\endgroup\@href {#1}{\urlprefix }}%
\providecommand \urlprefix  [0]{URL }%
\providecommand \Eprint [0]{\href }%
\providecommand \doibase [0]{https://doi.org/}%
\providecommand \selectlanguage [0]{\@gobble}%
\providecommand \bibinfo  [0]{\@secondoftwo}%
\providecommand \bibfield  [0]{\@secondoftwo}%
\providecommand \translation [1]{[#1]}%
\providecommand \BibitemOpen [0]{}%
\providecommand \bibitemStop [0]{}%
\providecommand \bibitemNoStop [0]{.\EOS\space}%
\providecommand \EOS [0]{\spacefactor3000\relax}%
\providecommand \BibitemShut  [1]{\csname bibitem#1\endcsname}%
\let\auto@bib@innerbib\@empty
%</preamble>
\bibitem [{\citenamefont {{Burrows}}\ and\ \citenamefont
  {{Vartanyan}}(2021)}]{Burrows2021}%
  \BibitemOpen
  \bibfield  {author} {\bibinfo {author} {\bibfnamefont {A.}~\bibnamefont
  {{Burrows}}}\ and\ \bibinfo {author} {\bibfnamefont {D.}~\bibnamefont
  {{Vartanyan}}},\ }\bibfield  {title} {\bibinfo {title} {{Core-collapse
  supernova explosion theory}},\ }\href
  {https://doi.org/10.1038/s41586-020-03059-w} {\bibfield  {journal} {\bibinfo
  {journal} {\nat}\ }\textbf {\bibinfo {volume} {589}},\ \bibinfo {pages} {29}
  (\bibinfo {year} {2021})},\ \Eprint {https://arxiv.org/abs/2009.14157}
  {arXiv:2009.14157 [astro-ph.SR]} \BibitemShut {NoStop}%
\bibitem [{\citenamefont {{Kistler}}\ \emph {et~al.}(2013)\citenamefont
  {{Kistler}}, \citenamefont {{Haxton}},\ and\ \citenamefont
  {{Y{\"u}ksel}}}]{Kistler2013}%
  \BibitemOpen
  \bibfield  {author} {\bibinfo {author} {\bibfnamefont {M.~D.}\ \bibnamefont
  {{Kistler}}}, \bibinfo {author} {\bibfnamefont {W.~C.}\ \bibnamefont
  {{Haxton}}},\ and\ \bibinfo {author} {\bibfnamefont {H.}~\bibnamefont
  {{Y{\"u}ksel}}},\ }\bibfield  {title} {\bibinfo {title} {{Tomography of
  Massive Stars from Core Collapse to Supernova Shock Breakout}},\ }\href
  {https://doi.org/10.1088/0004-637X/778/1/81} {\bibfield  {journal} {\bibinfo
  {journal} {\apj}\ }\textbf {\bibinfo {volume} {778}},\ \bibinfo {eid} {81}
  (\bibinfo {year} {2013})},\ \Eprint {https://arxiv.org/abs/1211.6770}
  {arXiv:1211.6770 [astro-ph.CO]} \BibitemShut {NoStop}%
\bibitem [{\citenamefont {{Janka}}(2012)}]{Janka2012}%
  \BibitemOpen
  \bibfield  {author} {\bibinfo {author} {\bibfnamefont {H.-T.}\ \bibnamefont
  {{Janka}}},\ }\bibfield  {title} {\bibinfo {title} {{Explosion Mechanisms of
  Core-Collapse Supernovae}},\ }\href
  {https://doi.org/10.1146/annurev-nucl-102711-094901} {\bibfield  {journal}
  {\bibinfo  {journal} {Annual Review of Nuclear and Particle Science}\
  }\textbf {\bibinfo {volume} {62}},\ \bibinfo {pages} {407} (\bibinfo {year}
  {2012})},\ \Eprint {https://arxiv.org/abs/1206.2503} {arXiv:1206.2503
  [astro-ph.SR]} \BibitemShut {NoStop}%
\bibitem [{\citenamefont {{M{\"u}ller}}(2019)}]{Muller2019}%
  \BibitemOpen
  \bibfield  {author} {\bibinfo {author} {\bibfnamefont {B.}~\bibnamefont
  {{M{\"u}ller}}},\ }\bibfield  {title} {\bibinfo {title} {{Neutrino Emission
  as Diagnostics of Core-Collapse Supernovae}},\ }\href
  {https://doi.org/10.1146/annurev-nucl-101918-023434} {\bibfield  {journal}
  {\bibinfo  {journal} {Annual Review of Nuclear and Particle Science}\
  }\textbf {\bibinfo {volume} {69}},\ \bibinfo {pages} {253} (\bibinfo {year}
  {2019})},\ \Eprint {https://arxiv.org/abs/1904.11067} {arXiv:1904.11067
  [astro-ph.HE]} \BibitemShut {NoStop}%
\bibitem [{\citenamefont {{Abdikamalov}}\ \emph {et~al.}(2020)\citenamefont
  {{Abdikamalov}}, \citenamefont {{Pagliaroli}},\ and\ \citenamefont
  {{Radice}}}]{Abdikamalov2020arXiv}%
  \BibitemOpen
  \bibfield  {author} {\bibinfo {author} {\bibfnamefont {E.}~\bibnamefont
  {{Abdikamalov}}}, \bibinfo {author} {\bibfnamefont {G.}~\bibnamefont
  {{Pagliaroli}}},\ and\ \bibinfo {author} {\bibfnamefont {D.}~\bibnamefont
  {{Radice}}},\ }\bibfield  {title} {\bibinfo {title} {{Gravitational Waves
  from Core-Collapse Supernovae}},\ }\href@noop {} {\bibfield  {journal}
  {\bibinfo  {journal} {arXiv e-prints}\ ,\ \bibinfo {eid} {arXiv:2010.04356}}
  (\bibinfo {year} {2020})},\ \Eprint {https://arxiv.org/abs/2010.04356}
  {arXiv:2010.04356 [astro-ph.SR]} \BibitemShut {NoStop}%
\bibitem [{\citenamefont {{Hirata}}\ \emph {et~al.}(1987)\citenamefont
  {{Hirata}} \emph {et~al.}}]{Hirata1987}%
  \BibitemOpen
  \bibfield  {author} {\bibinfo {author} {\bibfnamefont {K.}~\bibnamefont
  {{Hirata}}} \emph {et~al.},\ }\bibfield  {title} {\bibinfo {title}
  {{Observation of a neutrino burst from the supernova SN1987A}},\ }\href
  {https://doi.org/10.1103/PhysRevLett.58.1490} {\bibfield  {journal} {\bibinfo
   {journal} {\prl}\ }\textbf {\bibinfo {volume} {58}},\ \bibinfo {pages}
  {1490} (\bibinfo {year} {1987})}\BibitemShut {NoStop}%
\bibitem [{\citenamefont {{Bionta}}\ \emph {et~al.}(1987)\citenamefont
  {{Bionta}} \emph {et~al.}}]{Bionta1987}%
  \BibitemOpen
  \bibfield  {author} {\bibinfo {author} {\bibfnamefont {R.~M.}\ \bibnamefont
  {{Bionta}}} \emph {et~al.},\ }\bibfield  {title} {\bibinfo {title}
  {{Observation of a neutrino burst in coincidence with supernova 1987A in the
  Large Magellanic Cloud}},\ }\href
  {https://doi.org/10.1103/PhysRevLett.58.1494} {\bibfield  {journal} {\bibinfo
   {journal} {\prl}\ }\textbf {\bibinfo {volume} {58}},\ \bibinfo {pages}
  {1494} (\bibinfo {year} {1987})}\BibitemShut {NoStop}%
\bibitem [{\citenamefont {{Alexeyev}}\ \emph {et~al.}(1988)\citenamefont
  {{Alexeyev}}, \citenamefont {{Alexeyeva}}, \citenamefont {{Krivosheina}},\
  and\ \citenamefont {{Volchenko}}}]{Alexeyev1988}%
  \BibitemOpen
  \bibfield  {author} {\bibinfo {author} {\bibfnamefont {E.~N.}\ \bibnamefont
  {{Alexeyev}}}, \bibinfo {author} {\bibfnamefont {L.~N.}\ \bibnamefont
  {{Alexeyeva}}}, \bibinfo {author} {\bibfnamefont {I.~V.}\ \bibnamefont
  {{Krivosheina}}},\ and\ \bibinfo {author} {\bibfnamefont {V.~I.}\
  \bibnamefont {{Volchenko}}},\ }\bibfield  {title} {\bibinfo {title}
  {{Detection of the neutrino signal from SN 1987A in the LMC using the INR
  Baksan underground scintillation telescope}},\ }\href
  {https://doi.org/10.1016/0370-2693(88)91651-6} {\bibfield  {journal}
  {\bibinfo  {journal} {Physics Letters B}\ }\textbf {\bibinfo {volume}
  {205}},\ \bibinfo {pages} {209} (\bibinfo {year} {1988})}\BibitemShut
  {NoStop}%
\bibitem [{\citenamefont {{Abbott}}\ \emph
  {et~al.}(2020{\natexlab{a}})\citenamefont {{Abbott}} \emph
  {et~al.}}]{LVC2020}%
  \BibitemOpen
  \bibfield  {author} {\bibinfo {author} {\bibfnamefont {B.~P.}\ \bibnamefont
  {{Abbott}}} \emph {et~al.} (\bibinfo {collaboration} {{LIGO Scientific
  Collaboration} and {Virgo Collaboration}}),\ }\bibfield  {title} {\bibinfo
  {title} {{Optically targeted search for gravitational waves emitted by
  core-collapse supernovae during the first and second observing runs of
  advanced LIGO and advanced Virgo}},\ }\href
  {https://doi.org/10.1103/PhysRevD.101.084002} {\bibfield  {journal} {\bibinfo
   {journal} {\prd}\ }\textbf {\bibinfo {volume} {101}},\ \bibinfo {eid}
  {084002} (\bibinfo {year} {2020}{\natexlab{a}})},\ \Eprint
  {https://arxiv.org/abs/1908.03584} {arXiv:1908.03584 [astro-ph.HE]}
  \BibitemShut {NoStop}%
\bibitem [{\citenamefont {{Abbott}}\ \emph {et~al.}(2021)\citenamefont
  {{Abbott}} \emph {et~al.}}]{LVK2021arXiv}%
  \BibitemOpen
  \bibfield  {author} {\bibinfo {author} {\bibfnamefont {R.}~\bibnamefont
  {{Abbott}}} \emph {et~al.} (\bibinfo {collaboration} {{The LIGO Scientific
  Collaboration} and {the Virgo Collaboration} and {the KAGRA
  Collaboration}}),\ }\bibfield  {title} {\bibinfo {title} {{All-sky search for
  short gravitational-wave bursts in the third Advanced LIGO and Advanced Virgo
  run}},\ }\href@noop {} {\bibfield  {journal} {\bibinfo  {journal} {arXiv
  e-prints}\ ,\ \bibinfo {eid} {arXiv:2107.03701}} (\bibinfo {year} {2021})},\
  \Eprint {https://arxiv.org/abs/2107.03701} {arXiv:2107.03701 [gr-qc]}
  \BibitemShut {NoStop}%
\bibitem [{\citenamefont {{Rozwadowska}}\ \emph {et~al.}(2021)\citenamefont
  {{Rozwadowska}}, \citenamefont {{Vissani}},\ and\ \citenamefont
  {{Cappellaro}}}]{Rozwadowska2021}%
  \BibitemOpen
  \bibfield  {author} {\bibinfo {author} {\bibfnamefont {K.}~\bibnamefont
  {{Rozwadowska}}}, \bibinfo {author} {\bibfnamefont {F.}~\bibnamefont
  {{Vissani}}},\ and\ \bibinfo {author} {\bibfnamefont {E.}~\bibnamefont
  {{Cappellaro}}},\ }\bibfield  {title} {\bibinfo {title} {{On the rate of core
  collapse supernovae in the milky way}},\ }\href
  {https://doi.org/10.1016/j.newast.2020.101498} {\bibfield  {journal}
  {\bibinfo  {journal} {\na}\ }\textbf {\bibinfo {volume} {83}},\ \bibinfo
  {eid} {101498} (\bibinfo {year} {2021})},\ \Eprint
  {https://arxiv.org/abs/2009.03438} {arXiv:2009.03438 [astro-ph.HE]}
  \BibitemShut {NoStop}%
\bibitem [{\citenamefont {{Aasi}}\ \emph {et~al.}(2015)\citenamefont {{Aasi}}
  \emph {et~al.}}]{aLIGO2015}%
  \BibitemOpen
  \bibfield  {author} {\bibinfo {author} {\bibfnamefont {J.}~\bibnamefont
  {{Aasi}}} \emph {et~al.} (\bibinfo {collaboration} {LIGO Scientific
  Collaboration}),\ }\bibfield  {title} {\bibinfo {title} {{Advanced LIGO}},\
  }\href {https://doi.org/10.1088/0264-9381/32/7/074001} {\bibfield  {journal}
  {\bibinfo  {journal} {Classical and Quantum Gravity}\ }\textbf {\bibinfo
  {volume} {32}},\ \bibinfo {eid} {074001} (\bibinfo {year} {2015})},\ \Eprint
  {https://arxiv.org/abs/1411.4547} {arXiv:1411.4547 [gr-qc]} \BibitemShut
  {NoStop}%
\bibitem [{\citenamefont {{Acernese}}\ \emph {et~al.}(2015)\citenamefont
  {{Acernese}} \emph {et~al.}}]{AdVirgo2015}%
  \BibitemOpen
  \bibfield  {author} {\bibinfo {author} {\bibfnamefont {F.}~\bibnamefont
  {{Acernese}}} \emph {et~al.} (\bibinfo {collaboration} {Virgo
  Collaboration}),\ }\bibfield  {title} {\bibinfo {title} {{Advanced Virgo: a
  second-generation interferometric gravitational wave detector}},\ }\href
  {https://doi.org/10.1088/0264-9381/32/2/024001} {\bibfield  {journal}
  {\bibinfo  {journal} {Classical and Quantum Gravity}\ }\textbf {\bibinfo
  {volume} {32}},\ \bibinfo {eid} {024001} (\bibinfo {year} {2015})},\ \Eprint
  {https://arxiv.org/abs/1408.3978} {arXiv:1408.3978 [gr-qc]} \BibitemShut
  {NoStop}%
\bibitem [{\citenamefont {{Abbott}}\ \emph
  {et~al.}(2020{\natexlab{b}})\citenamefont {{Abbott}} \emph
  {et~al.}}]{LVK2020}%
  \BibitemOpen
  \bibfield  {author} {\bibinfo {author} {\bibfnamefont {B.~P.}\ \bibnamefont
  {{Abbott}}} \emph {et~al.} (\bibinfo {collaboration} {{Kagra Collaboration},
  {LIGO Scientific Collaboration} and {VIRGO Collaboration}}),\ }\bibfield
  {title} {\bibinfo {title} {{Prospects for observing and localizing
  gravitational-wave transients with Advanced LIGO, Advanced Virgo and
  KAGRA}},\ }\href {https://doi.org/10.1007/s41114-020-00026-9} {\bibfield
  {journal} {\bibinfo  {journal} {Living Reviews in Relativity}\ }\textbf
  {\bibinfo {volume} {23}},\ \bibinfo {eid} {3} (\bibinfo {year}
  {2020}{\natexlab{b}})},\ \Eprint {https://arxiv.org/abs/1304.0670}
  {arXiv:1304.0670 [gr-qc]} \BibitemShut {NoStop}%
\bibitem [{\citenamefont {{Somiya}}(2012)}]{Somiya2012}%
  \BibitemOpen
  \bibfield  {author} {\bibinfo {author} {\bibfnamefont {K.}~\bibnamefont
  {{Somiya}}},\ }\bibfield  {title} {\bibinfo {title} {{Detector configuration
  of KAGRA-the Japanese cryogenic gravitational-wave detector}},\ }\href
  {https://doi.org/10.1088/0264-9381/29/12/124007} {\bibfield  {journal}
  {\bibinfo  {journal} {Classical and Quantum Gravity}\ }\textbf {\bibinfo
  {volume} {29}},\ \bibinfo {eid} {124007} (\bibinfo {year} {2012})},\ \Eprint
  {https://arxiv.org/abs/1111.7185} {arXiv:1111.7185 [gr-qc]} \BibitemShut
  {NoStop}%
\bibitem [{\citenamefont {{Aso}}\ \emph {et~al.}(2013)\citenamefont {{Aso}},
  \citenamefont {{Michimura}}, \citenamefont {{Somiya}}, \citenamefont
  {{Ando}}, \citenamefont {{Miyakawa}}, \citenamefont {{Sekiguchi}},
  \citenamefont {{Tatsumi}},\ and\ \citenamefont {{Yamamoto}}}]{Aso2013}%
  \BibitemOpen
  \bibfield  {author} {\bibinfo {author} {\bibfnamefont {Y.}~\bibnamefont
  {{Aso}}}, \bibinfo {author} {\bibfnamefont {Y.}~\bibnamefont {{Michimura}}},
  \bibinfo {author} {\bibfnamefont {K.}~\bibnamefont {{Somiya}}}, \bibinfo
  {author} {\bibfnamefont {M.}~\bibnamefont {{Ando}}}, \bibinfo {author}
  {\bibfnamefont {O.}~\bibnamefont {{Miyakawa}}}, \bibinfo {author}
  {\bibfnamefont {T.}~\bibnamefont {{Sekiguchi}}}, \bibinfo {author}
  {\bibfnamefont {D.}~\bibnamefont {{Tatsumi}}},\ and\ \bibinfo {author}
  {\bibfnamefont {H.}~\bibnamefont {{Yamamoto}}},\ }\bibfield  {title}
  {\bibinfo {title} {{Interferometer design of the KAGRA gravitational wave
  detector}},\ }\href {https://doi.org/10.1103/PhysRevD.88.043007} {\bibfield
  {journal} {\bibinfo  {journal} {\prd}\ }\textbf {\bibinfo {volume} {88}},\
  \bibinfo {eid} {043007} (\bibinfo {year} {2013})},\ \Eprint
  {https://arxiv.org/abs/1306.6747} {arXiv:1306.6747 [gr-qc]} \BibitemShut
  {NoStop}%
\bibitem [{\citenamefont {{Iyer}}\ \emph {et~al.}(2011)\citenamefont {{Iyer}}
  \emph {et~al.}}]{Iyer2011}%
  \BibitemOpen
  \bibfield  {author} {\bibinfo {author} {\bibfnamefont {B.}~\bibnamefont
  {{Iyer}}} \emph {et~al.},\ }\bibfield  {title} {\bibinfo {title}
  {{LIGO-India. Technical Report M1100296-v2}},\ }\href
  {https://dcc.ligo.org/LIGO-M1100296/public} {\bibfield  {journal} {\bibinfo
  {journal} {IndIGO, India}\ } (\bibinfo {year} {2011})}\BibitemShut {NoStop}%
\bibitem [{\citenamefont {{Gossan}}\ \emph {et~al.}(2016)\citenamefont
  {{Gossan}}, \citenamefont {{Sutton}}, \citenamefont {{Stuver}}, \citenamefont
  {{Zanolin}}, \citenamefont {{Gill}},\ and\ \citenamefont
  {{Ott}}}]{Gossan2016}%
  \BibitemOpen
  \bibfield  {author} {\bibinfo {author} {\bibfnamefont {S.~E.}\ \bibnamefont
  {{Gossan}}}, \bibinfo {author} {\bibfnamefont {P.}~\bibnamefont {{Sutton}}},
  \bibinfo {author} {\bibfnamefont {A.}~\bibnamefont {{Stuver}}}, \bibinfo
  {author} {\bibfnamefont {M.}~\bibnamefont {{Zanolin}}}, \bibinfo {author}
  {\bibfnamefont {K.}~\bibnamefont {{Gill}}},\ and\ \bibinfo {author}
  {\bibfnamefont {C.~D.}\ \bibnamefont {{Ott}}},\ }\bibfield  {title} {\bibinfo
  {title} {{Observing gravitational waves from core-collapse supernovae in the
  advanced detector era}},\ }\href {https://doi.org/10.1103/PhysRevD.93.042002}
  {\bibfield  {journal} {\bibinfo  {journal} {\prd}\ }\textbf {\bibinfo
  {volume} {93}},\ \bibinfo {eid} {042002} (\bibinfo {year} {2016})},\ \Eprint
  {https://arxiv.org/abs/1511.02836} {arXiv:1511.02836 [astro-ph.HE]}
  \BibitemShut {NoStop}%
\bibitem [{\citenamefont {{Szczepa{\'n}czyk}}\ \emph
  {et~al.}(2021)\citenamefont {{Szczepa{\'n}czyk}} \emph
  {et~al.}}]{Szczepanczyk2021}%
  \BibitemOpen
  \bibfield  {author} {\bibinfo {author} {\bibfnamefont {M.~J.}\ \bibnamefont
  {{Szczepa{\'n}czyk}}} \emph {et~al.},\ }\bibfield  {title} {\bibinfo {title}
  {{Detecting and reconstructing gravitational waves from the next galactic
  core-collapse supernova in the advanced detector era}},\ }\href
  {https://doi.org/10.1103/PhysRevD.104.102002} {\bibfield  {journal} {\bibinfo
   {journal} {\prd}\ }\textbf {\bibinfo {volume} {104}},\ \bibinfo {eid}
  {102002} (\bibinfo {year} {2021})},\ \Eprint
  {https://arxiv.org/abs/2104.06462} {arXiv:2104.06462 [astro-ph.HE]}
  \BibitemShut {NoStop}%
\bibitem [{\citenamefont {{Cornish}}\ and\ \citenamefont
  {{Littenberg}}(2015)}]{Cornish2015}%
  \BibitemOpen
  \bibfield  {author} {\bibinfo {author} {\bibfnamefont {N.~J.}\ \bibnamefont
  {{Cornish}}}\ and\ \bibinfo {author} {\bibfnamefont {T.~B.}\ \bibnamefont
  {{Littenberg}}},\ }\bibfield  {title} {\bibinfo {title} {{Bayeswave: Bayesian
  inference for gravitational wave bursts and instrument glitches}},\ }\href
  {https://doi.org/10.1088/0264-9381/32/13/135012} {\bibfield  {journal}
  {\bibinfo  {journal} {Classical and Quantum Gravity}\ }\textbf {\bibinfo
  {volume} {32}},\ \bibinfo {eid} {135012} (\bibinfo {year} {2015})},\ \Eprint
  {https://arxiv.org/abs/1410.3835} {arXiv:1410.3835 [gr-qc]} \BibitemShut
  {NoStop}%
\bibitem [{\citenamefont {{Woosley}}\ \emph {et~al.}(2002)\citenamefont
  {{Woosley}}, \citenamefont {{Heger}},\ and\ \citenamefont
  {{Weaver}}}]{Woosley2002}%
  \BibitemOpen
  \bibfield  {author} {\bibinfo {author} {\bibfnamefont {S.~E.}\ \bibnamefont
  {{Woosley}}}, \bibinfo {author} {\bibfnamefont {A.}~\bibnamefont {{Heger}}},\
  and\ \bibinfo {author} {\bibfnamefont {T.~A.}\ \bibnamefont {{Weaver}}},\
  }\bibfield  {title} {\bibinfo {title} {{The evolution and explosion of
  massive stars}},\ }\href {https://doi.org/10.1103/RevModPhys.74.1015}
  {\bibfield  {journal} {\bibinfo  {journal} {Reviews of Modern Physics}\
  }\textbf {\bibinfo {volume} {74}},\ \bibinfo {pages} {1015} (\bibinfo {year}
  {2002})}\BibitemShut {NoStop}%
\bibitem [{\citenamefont {{Janka}}(2017)}]{Janka2017}%
  \BibitemOpen
  \bibfield  {author} {\bibinfo {author} {\bibfnamefont {H.-T.}\ \bibnamefont
  {{Janka}}},\ }\bibinfo {title} {{Neutrino-Driven Explosions}},\ in\ \href
  {https://doi.org/10.1007/978-3-319-21846-5\_109} {\emph {\bibinfo {booktitle}
  {Handbook of Supernovae}}},\ \bibinfo {editor} {edited by\ \bibinfo {editor}
  {\bibfnamefont {A.~W.}\ \bibnamefont {{Alsabti}}}\ and\ \bibinfo {editor}
  {\bibfnamefont {P.}~\bibnamefont {{Murdin}}}}\ (\bibinfo  {publisher}
  {Springer International Publishing},\ \bibinfo {year} {2017})\ p.\ \bibinfo
  {pages} {1095}\BibitemShut {NoStop}%
\bibitem [{\citenamefont {{Woosley}}\ and\ \citenamefont
  {{Heger}}(2006)}]{Woosley2006}%
  \BibitemOpen
  \bibfield  {author} {\bibinfo {author} {\bibfnamefont {S.~E.}\ \bibnamefont
  {{Woosley}}}\ and\ \bibinfo {author} {\bibfnamefont {A.}~\bibnamefont
  {{Heger}}},\ }\bibfield  {title} {\bibinfo {title} {{The Progenitor Stars of
  Gamma-Ray Bursts}},\ }\href {https://doi.org/10.1086/498500} {\bibfield
  {journal} {\bibinfo  {journal} {\apj}\ }\textbf {\bibinfo {volume} {637}},\
  \bibinfo {pages} {914} (\bibinfo {year} {2006})},\ \Eprint
  {https://arxiv.org/abs/astro-ph/0508175} {arXiv:astro-ph/0508175 [astro-ph]}
  \BibitemShut {NoStop}%
\bibitem [{\citenamefont {{Obergaulinger}}\ and\ \citenamefont
  {{Aloy}}(2020)}]{Obergaulinger2020}%
  \BibitemOpen
  \bibfield  {author} {\bibinfo {author} {\bibfnamefont {M.}~\bibnamefont
  {{Obergaulinger}}}\ and\ \bibinfo {author} {\bibfnamefont {M.~{\'A}.}\
  \bibnamefont {{Aloy}}},\ }\bibfield  {title} {\bibinfo {title}
  {{Magnetorotational core collapse of possible GRB progenitors - I. Explosion
  mechanisms}},\ }\href {https://doi.org/10.1093/mnras/staa096} {\bibfield
  {journal} {\bibinfo  {journal} {\mnras}\ }\textbf {\bibinfo {volume} {492}},\
  \bibinfo {pages} {4613} (\bibinfo {year} {2020})},\ \Eprint
  {https://arxiv.org/abs/1909.01105} {arXiv:1909.01105 [astro-ph.HE]}
  \BibitemShut {NoStop}%
\bibitem [{\citenamefont {{Torres-Forn{\'e}}}\ \emph
  {et~al.}(2018)\citenamefont {{Torres-Forn{\'e}}}, \citenamefont
  {{Cerd{\'a}-Dur{\'a}n}}, \citenamefont {{Passamonti}},\ and\ \citenamefont
  {{Font}}}]{Torres-Forne2018}%
  \BibitemOpen
  \bibfield  {author} {\bibinfo {author} {\bibfnamefont {A.}~\bibnamefont
  {{Torres-Forn{\'e}}}}, \bibinfo {author} {\bibfnamefont {P.}~\bibnamefont
  {{Cerd{\'a}-Dur{\'a}n}}}, \bibinfo {author} {\bibfnamefont {A.}~\bibnamefont
  {{Passamonti}}},\ and\ \bibinfo {author} {\bibfnamefont {J.~A.}\ \bibnamefont
  {{Font}}},\ }\bibfield  {title} {\bibinfo {title} {{Towards asteroseismology
  of core-collapse supernovae with gravitational-wave observations - I. Cowling
  approximation}},\ }\href {https://doi.org/10.1093/mnras/stx3067} {\bibfield
  {journal} {\bibinfo  {journal} {\mnras}\ }\textbf {\bibinfo {volume} {474}},\
  \bibinfo {pages} {5272} (\bibinfo {year} {2018})},\ \Eprint
  {https://arxiv.org/abs/1708.01920} {arXiv:1708.01920 [astro-ph.SR]}
  \BibitemShut {NoStop}%
\bibitem [{\citenamefont {{O'Connor}}\ and\ \citenamefont
  {{Couch}}(2018)}]{OConnor2018}%
  \BibitemOpen
  \bibfield  {author} {\bibinfo {author} {\bibfnamefont {E.~P.}\ \bibnamefont
  {{O'Connor}}}\ and\ \bibinfo {author} {\bibfnamefont {S.~M.}\ \bibnamefont
  {{Couch}}},\ }\bibfield  {title} {\bibinfo {title} {{Exploring Fundamentally
  Three-dimensional Phenomena in High-fidelity Simulations of Core-collapse
  Supernovae}},\ }\href {https://doi.org/10.3847/1538-4357/aadcf7} {\bibfield
  {journal} {\bibinfo  {journal} {\apj}\ }\textbf {\bibinfo {volume} {865}},\
  \bibinfo {eid} {81} (\bibinfo {year} {2018})},\ \Eprint
  {https://arxiv.org/abs/1807.07579} {arXiv:1807.07579 [astro-ph.HE]}
  \BibitemShut {NoStop}%
\bibitem [{\citenamefont {{Radice}}\ \emph {et~al.}(2019)\citenamefont
  {{Radice}}, \citenamefont {{Morozova}}, \citenamefont {{Burrows}},
  \citenamefont {{Vartanyan}},\ and\ \citenamefont {{Nagakura}}}]{Radice2019}%
  \BibitemOpen
  \bibfield  {author} {\bibinfo {author} {\bibfnamefont {D.}~\bibnamefont
  {{Radice}}}, \bibinfo {author} {\bibfnamefont {V.}~\bibnamefont
  {{Morozova}}}, \bibinfo {author} {\bibfnamefont {A.}~\bibnamefont
  {{Burrows}}}, \bibinfo {author} {\bibfnamefont {D.}~\bibnamefont
  {{Vartanyan}}},\ and\ \bibinfo {author} {\bibfnamefont {H.}~\bibnamefont
  {{Nagakura}}},\ }\bibfield  {title} {\bibinfo {title} {{Characterizing the
  Gravitational Wave Signal from Core-collapse Supernovae}},\ }\href
  {https://doi.org/10.3847/2041-8213/ab191a} {\bibfield  {journal} {\bibinfo
  {journal} {\apjl}\ }\textbf {\bibinfo {volume} {876}},\ \bibinfo {eid} {L9}
  (\bibinfo {year} {2019})},\ \Eprint {https://arxiv.org/abs/1812.07703}
  {arXiv:1812.07703 [astro-ph.HE]} \BibitemShut {NoStop}%
\bibitem [{\citenamefont {{Powell}}\ and\ \citenamefont
  {{M{\"u}ller}}(2019)}]{Powell2019}%
  \BibitemOpen
  \bibfield  {author} {\bibinfo {author} {\bibfnamefont {J.}~\bibnamefont
  {{Powell}}}\ and\ \bibinfo {author} {\bibfnamefont {B.}~\bibnamefont
  {{M{\"u}ller}}},\ }\bibfield  {title} {\bibinfo {title} {{Gravitational wave
  emission from 3D explosion models of core-collapse supernovae with low and
  normal explosion energies}},\ }\href {https://doi.org/10.1093/mnras/stz1304}
  {\bibfield  {journal} {\bibinfo  {journal} {\mnras}\ }\textbf {\bibinfo
  {volume} {487}},\ \bibinfo {pages} {1178} (\bibinfo {year} {2019})},\ \Eprint
  {https://arxiv.org/abs/1812.05738} {arXiv:1812.05738 [astro-ph.HE]}
  \BibitemShut {NoStop}%
\bibitem [{\citenamefont {{Torres-Forn{\'e}}}\ \emph
  {et~al.}(2019)\citenamefont {{Torres-Forn{\'e}}}, \citenamefont
  {{Cerd{\'a}-Dur{\'a}n}}, \citenamefont {{Obergaulinger}}, \citenamefont
  {{M{\"u}ller}},\ and\ \citenamefont {{Font}}}]{Torres-Forne2019}%
  \BibitemOpen
  \bibfield  {author} {\bibinfo {author} {\bibfnamefont {A.}~\bibnamefont
  {{Torres-Forn{\'e}}}}, \bibinfo {author} {\bibfnamefont {P.}~\bibnamefont
  {{Cerd{\'a}-Dur{\'a}n}}}, \bibinfo {author} {\bibfnamefont {M.}~\bibnamefont
  {{Obergaulinger}}}, \bibinfo {author} {\bibfnamefont {B.}~\bibnamefont
  {{M{\"u}ller}}},\ and\ \bibinfo {author} {\bibfnamefont {J.~A.}\ \bibnamefont
  {{Font}}},\ }\bibfield  {title} {\bibinfo {title} {{Universal Relations for
  Gravitational-Wave Asteroseismology of Protoneutron Stars}},\ }\href
  {https://doi.org/10.1103/PhysRevLett.123.051102} {\bibfield  {journal}
  {\bibinfo  {journal} {\prl}\ }\textbf {\bibinfo {volume} {123}},\ \bibinfo
  {eid} {051102} (\bibinfo {year} {2019})},\ \Eprint
  {https://arxiv.org/abs/1902.10048} {arXiv:1902.10048 [gr-qc]} \BibitemShut
  {NoStop}%
\bibitem [{\citenamefont {{Bizouard}}\ \emph {et~al.}(2021)\citenamefont
  {{Bizouard}}, \citenamefont {{Maturana-Russel}}, \citenamefont
  {{Torres-Forn{\'e}}}, \citenamefont {{Obergaulinger}}, \citenamefont
  {{Cerd{\'a}-Dur{\'a}n}}, \citenamefont {{Christensen}}, \citenamefont
  {{Font}},\ and\ \citenamefont {{Meyer}}}]{Bizouard2021}%
  \BibitemOpen
  \bibfield  {author} {\bibinfo {author} {\bibfnamefont {M.-A.}\ \bibnamefont
  {{Bizouard}}}, \bibinfo {author} {\bibfnamefont {P.}~\bibnamefont
  {{Maturana-Russel}}}, \bibinfo {author} {\bibfnamefont {A.}~\bibnamefont
  {{Torres-Forn{\'e}}}}, \bibinfo {author} {\bibfnamefont {M.}~\bibnamefont
  {{Obergaulinger}}}, \bibinfo {author} {\bibfnamefont {P.}~\bibnamefont
  {{Cerd{\'a}-Dur{\'a}n}}}, \bibinfo {author} {\bibfnamefont {N.}~\bibnamefont
  {{Christensen}}}, \bibinfo {author} {\bibfnamefont {J.~A.}\ \bibnamefont
  {{Font}}},\ and\ \bibinfo {author} {\bibfnamefont {R.}~\bibnamefont
  {{Meyer}}},\ }\bibfield  {title} {\bibinfo {title} {{Inference of
  protoneutron star properties from gravitational-wave data in core-collapse
  supernovae}},\ }\href {https://doi.org/10.1103/PhysRevD.103.063006}
  {\bibfield  {journal} {\bibinfo  {journal} {\prd}\ }\textbf {\bibinfo
  {volume} {103}},\ \bibinfo {eid} {063006} (\bibinfo {year} {2021})},\ \Eprint
  {https://arxiv.org/abs/2012.00846} {arXiv:2012.00846 [gr-qc]} \BibitemShut
  {NoStop}%
\bibitem [{\citenamefont {{Sotani}}\ \emph {et~al.}(2021)\citenamefont
  {{Sotani}}, \citenamefont {{Takiwaki}},\ and\ \citenamefont
  {{Togashi}}}]{Sotani2021}%
  \BibitemOpen
  \bibfield  {author} {\bibinfo {author} {\bibfnamefont {H.}~\bibnamefont
  {{Sotani}}}, \bibinfo {author} {\bibfnamefont {T.}~\bibnamefont
  {{Takiwaki}}},\ and\ \bibinfo {author} {\bibfnamefont {H.}~\bibnamefont
  {{Togashi}}},\ }\bibfield  {title} {\bibinfo {title} {{Universal relation for
  supernova gravitational waves}},\ }\href
  {https://doi.org/10.1103/PhysRevD.104.123009} {\bibfield  {journal} {\bibinfo
   {journal} {\prd}\ }\textbf {\bibinfo {volume} {104}},\ \bibinfo {eid}
  {123009} (\bibinfo {year} {2021})},\ \Eprint
  {https://arxiv.org/abs/2110.03131} {arXiv:2110.03131 [astro-ph.HE]}
  \BibitemShut {NoStop}%
\bibitem [{\citenamefont {{Blondin}}\ \emph {et~al.}(2003)\citenamefont
  {{Blondin}}, \citenamefont {{Mezzacappa}},\ and\ \citenamefont
  {{DeMarino}}}]{Blondin2003}%
  \BibitemOpen
  \bibfield  {author} {\bibinfo {author} {\bibfnamefont {J.~M.}\ \bibnamefont
  {{Blondin}}}, \bibinfo {author} {\bibfnamefont {A.}~\bibnamefont
  {{Mezzacappa}}},\ and\ \bibinfo {author} {\bibfnamefont {C.}~\bibnamefont
  {{DeMarino}}},\ }\bibfield  {title} {\bibinfo {title} {{Stability of Standing
  Accretion Shocks, with an Eye toward Core-Collapse Supernovae}},\ }\href
  {https://doi.org/10.1086/345812} {\bibfield  {journal} {\bibinfo  {journal}
  {\apj}\ }\textbf {\bibinfo {volume} {584}},\ \bibinfo {pages} {971} (\bibinfo
  {year} {2003})},\ \Eprint {https://arxiv.org/abs/astro-ph/0210634}
  {arXiv:astro-ph/0210634 [astro-ph]} \BibitemShut {NoStop}%
\bibitem [{\citenamefont {{Takiwaki}}\ and\ \citenamefont
  {{Kotake}}(2011)}]{Takiwaki2011}%
  \BibitemOpen
  \bibfield  {author} {\bibinfo {author} {\bibfnamefont {T.}~\bibnamefont
  {{Takiwaki}}}\ and\ \bibinfo {author} {\bibfnamefont {K.}~\bibnamefont
  {{Kotake}}},\ }\bibfield  {title} {\bibinfo {title} {{Gravitational Wave
  Signatures of Magnetohydrodynamically Driven Core-collapse Supernova
  Explosions}},\ }\href {https://doi.org/10.1088/0004-637X/743/1/30} {\bibfield
   {journal} {\bibinfo  {journal} {\apj}\ }\textbf {\bibinfo {volume} {743}},\
  \bibinfo {eid} {30} (\bibinfo {year} {2011})},\ \Eprint
  {https://arxiv.org/abs/1004.2896} {arXiv:1004.2896 [astro-ph.HE]}
  \BibitemShut {NoStop}%
\bibitem [{\citenamefont {{Richers}}\ \emph {et~al.}(2017)\citenamefont
  {{Richers}}, \citenamefont {{Ott}}, \citenamefont {{Abdikamalov}},
  \citenamefont {{O'Connor}},\ and\ \citenamefont {{Sullivan}}}]{Richers2017}%
  \BibitemOpen
  \bibfield  {author} {\bibinfo {author} {\bibfnamefont {S.}~\bibnamefont
  {{Richers}}}, \bibinfo {author} {\bibfnamefont {C.~D.}\ \bibnamefont
  {{Ott}}}, \bibinfo {author} {\bibfnamefont {E.}~\bibnamefont
  {{Abdikamalov}}}, \bibinfo {author} {\bibfnamefont {E.}~\bibnamefont
  {{O'Connor}}},\ and\ \bibinfo {author} {\bibfnamefont {C.}~\bibnamefont
  {{Sullivan}}},\ }\bibfield  {title} {\bibinfo {title} {{Equation of state
  effects on gravitational waves from rotating core collapse}},\ }\href
  {https://doi.org/10.1103/PhysRevD.95.063019} {\bibfield  {journal} {\bibinfo
  {journal} {\prd}\ }\textbf {\bibinfo {volume} {95}},\ \bibinfo {eid} {063019}
  (\bibinfo {year} {2017})},\ \Eprint {https://arxiv.org/abs/1701.02752}
  {arXiv:1701.02752 [astro-ph.HE]} \BibitemShut {NoStop}%
\bibitem [{\citenamefont {{Abdikamalov}}\ \emph {et~al.}(2014)\citenamefont
  {{Abdikamalov}}, \citenamefont {{Gossan}}, \citenamefont {{DeMaio}},\ and\
  \citenamefont {{Ott}}}]{Abdikamalov2014}%
  \BibitemOpen
  \bibfield  {author} {\bibinfo {author} {\bibfnamefont {E.}~\bibnamefont
  {{Abdikamalov}}}, \bibinfo {author} {\bibfnamefont {S.}~\bibnamefont
  {{Gossan}}}, \bibinfo {author} {\bibfnamefont {A.~M.}\ \bibnamefont
  {{DeMaio}}},\ and\ \bibinfo {author} {\bibfnamefont {C.~D.}\ \bibnamefont
  {{Ott}}},\ }\bibfield  {title} {\bibinfo {title} {{Measuring the angular
  momentum distribution in core-collapse supernova progenitors with
  gravitational waves}},\ }\href {https://doi.org/10.1103/PhysRevD.90.044001}
  {\bibfield  {journal} {\bibinfo  {journal} {\prd}\ }\textbf {\bibinfo
  {volume} {90}},\ \bibinfo {eid} {044001} (\bibinfo {year} {2014})},\ \Eprint
  {https://arxiv.org/abs/1311.3678} {arXiv:1311.3678 [astro-ph.SR]}
  \BibitemShut {NoStop}%
\bibitem [{\citenamefont {{Fuller}}\ \emph {et~al.}(2015)\citenamefont
  {{Fuller}}, \citenamefont {{Klion}}, \citenamefont {{Abdikamalov}},\ and\
  \citenamefont {{Ott}}}]{Fuller2015}%
  \BibitemOpen
  \bibfield  {author} {\bibinfo {author} {\bibfnamefont {J.}~\bibnamefont
  {{Fuller}}}, \bibinfo {author} {\bibfnamefont {H.}~\bibnamefont {{Klion}}},
  \bibinfo {author} {\bibfnamefont {E.}~\bibnamefont {{Abdikamalov}}},\ and\
  \bibinfo {author} {\bibfnamefont {C.~D.}\ \bibnamefont {{Ott}}},\ }\bibfield
  {title} {\bibinfo {title} {{Supernova seismology: gravitational wave
  signatures of rapidly rotating core collapse}},\ }\href
  {https://doi.org/10.1093/mnras/stv698} {\bibfield  {journal} {\bibinfo
  {journal} {\mnras}\ }\textbf {\bibinfo {volume} {450}},\ \bibinfo {pages}
  {414} (\bibinfo {year} {2015})},\ \Eprint {https://arxiv.org/abs/1501.06951}
  {arXiv:1501.06951 [astro-ph.HE]} \BibitemShut {NoStop}%
\bibitem [{\citenamefont {{Logue}}\ \emph {et~al.}(2012)\citenamefont
  {{Logue}}, \citenamefont {{Ott}}, \citenamefont {{Heng}}, \citenamefont
  {{Kalmus}},\ and\ \citenamefont {{Scargill}}}]{Logue2012}%
  \BibitemOpen
  \bibfield  {author} {\bibinfo {author} {\bibfnamefont {J.}~\bibnamefont
  {{Logue}}}, \bibinfo {author} {\bibfnamefont {C.~D.}\ \bibnamefont {{Ott}}},
  \bibinfo {author} {\bibfnamefont {I.~S.}\ \bibnamefont {{Heng}}}, \bibinfo
  {author} {\bibfnamefont {P.}~\bibnamefont {{Kalmus}}},\ and\ \bibinfo
  {author} {\bibfnamefont {J.~H.~C.}\ \bibnamefont {{Scargill}}},\ }\bibfield
  {title} {\bibinfo {title} {{Inferring core-collapse supernova physics with
  gravitational waves}},\ }\href {https://doi.org/10.1103/PhysRevD.86.044023}
  {\bibfield  {journal} {\bibinfo  {journal} {\prd}\ }\textbf {\bibinfo
  {volume} {86}},\ \bibinfo {eid} {044023} (\bibinfo {year} {2012})},\ \Eprint
  {https://arxiv.org/abs/1202.3256} {arXiv:1202.3256 [gr-qc]} \BibitemShut
  {NoStop}%
\bibitem [{\citenamefont {{Roma}}\ \emph {et~al.}(2019)\citenamefont {{Roma}},
  \citenamefont {{Powell}}, \citenamefont {{Heng}},\ and\ \citenamefont
  {{Frey}}}]{Roma2019}%
  \BibitemOpen
  \bibfield  {author} {\bibinfo {author} {\bibfnamefont {V.}~\bibnamefont
  {{Roma}}}, \bibinfo {author} {\bibfnamefont {J.}~\bibnamefont {{Powell}}},
  \bibinfo {author} {\bibfnamefont {I.~S.}\ \bibnamefont {{Heng}}},\ and\
  \bibinfo {author} {\bibfnamefont {R.}~\bibnamefont {{Frey}}},\ }\bibfield
  {title} {\bibinfo {title} {{Astrophysics with core-collapse supernova
  gravitational wave signals in the next generation of gravitational wave
  detectors}},\ }\href {https://doi.org/10.1103/PhysRevD.99.063018} {\bibfield
  {journal} {\bibinfo  {journal} {\prd}\ }\textbf {\bibinfo {volume} {99}},\
  \bibinfo {eid} {063018} (\bibinfo {year} {2019})},\ \Eprint
  {https://arxiv.org/abs/1901.08692} {arXiv:1901.08692 [astro-ph.IM]}
  \BibitemShut {NoStop}%
\bibitem [{\citenamefont {{Sutton}}\ \emph {et~al.}(2010)\citenamefont
  {{Sutton}} \emph {et~al.}}]{Sutton2010}%
  \BibitemOpen
  \bibfield  {author} {\bibinfo {author} {\bibfnamefont {P.~J.}\ \bibnamefont
  {{Sutton}}} \emph {et~al.},\ }\bibfield  {title} {\bibinfo {title}
  {{X-Pipeline: an analysis package for autonomous gravitational-wave burst
  searches}},\ }\href {https://doi.org/10.1088/1367-2630/12/5/053034}
  {\bibfield  {journal} {\bibinfo  {journal} {New Journal of Physics}\ }\textbf
  {\bibinfo {volume} {12}},\ \bibinfo {eid} {053034} (\bibinfo {year}
  {2010})},\ \Eprint {https://arxiv.org/abs/0908.3665} {arXiv:0908.3665
  [gr-qc]} \BibitemShut {NoStop}%
\bibitem [{\citenamefont {{Klimenko}}\ \emph {et~al.}(2016)\citenamefont
  {{Klimenko}} \emph {et~al.}}]{Klimenko2016}%
  \BibitemOpen
  \bibfield  {author} {\bibinfo {author} {\bibfnamefont {S.}~\bibnamefont
  {{Klimenko}}} \emph {et~al.},\ }\bibfield  {title} {\bibinfo {title} {{Method
  for detection and reconstruction of gravitational wave transients with
  networks of advanced detectors}},\ }\href
  {https://doi.org/10.1103/PhysRevD.93.042004} {\bibfield  {journal} {\bibinfo
  {journal} {\prd}\ }\textbf {\bibinfo {volume} {93}},\ \bibinfo {eid} {042004}
  (\bibinfo {year} {2016})},\ \Eprint {https://arxiv.org/abs/1511.05999}
  {arXiv:1511.05999 [gr-qc]} \BibitemShut {NoStop}%
\bibitem [{\citenamefont {{Adams}}\ \emph {et~al.}(2016)\citenamefont
  {{Adams}}, \citenamefont {{Buskulic}}, \citenamefont {{Germain}},
  \citenamefont {{Guidi}}, \citenamefont {{Marion}}, \citenamefont {{Montani}},
  \citenamefont {{Mours}}, \citenamefont {{Piergiovanni}},\ and\ \citenamefont
  {{Wang}}}]{Adams2016}%
  \BibitemOpen
  \bibfield  {author} {\bibinfo {author} {\bibfnamefont {T.}~\bibnamefont
  {{Adams}}}, \bibinfo {author} {\bibfnamefont {D.}~\bibnamefont {{Buskulic}}},
  \bibinfo {author} {\bibfnamefont {V.}~\bibnamefont {{Germain}}}, \bibinfo
  {author} {\bibfnamefont {G.~M.}\ \bibnamefont {{Guidi}}}, \bibinfo {author}
  {\bibfnamefont {F.}~\bibnamefont {{Marion}}}, \bibinfo {author}
  {\bibfnamefont {M.}~\bibnamefont {{Montani}}}, \bibinfo {author}
  {\bibfnamefont {B.}~\bibnamefont {{Mours}}}, \bibinfo {author} {\bibfnamefont
  {F.}~\bibnamefont {{Piergiovanni}}},\ and\ \bibinfo {author} {\bibfnamefont
  {G.}~\bibnamefont {{Wang}}},\ }\bibfield  {title} {\bibinfo {title}
  {{Low-latency analysis pipeline for compact binary coalescences in the
  advanced gravitational wave detector era}},\ }\href
  {https://doi.org/10.1088/0264-9381/33/17/175012} {\bibfield  {journal}
  {\bibinfo  {journal} {Classical and Quantum Gravity}\ }\textbf {\bibinfo
  {volume} {33}},\ \bibinfo {eid} {175012} (\bibinfo {year} {2016})},\ \Eprint
  {https://arxiv.org/abs/1512.02864} {arXiv:1512.02864 [gr-qc]} \BibitemShut
  {NoStop}%
\bibitem [{\citenamefont {{Usman}}\ \emph {et~al.}(2016)\citenamefont {{Usman}}
  \emph {et~al.}}]{Usman2016}%
  \BibitemOpen
  \bibfield  {author} {\bibinfo {author} {\bibfnamefont {S.~A.}\ \bibnamefont
  {{Usman}}} \emph {et~al.},\ }\bibfield  {title} {\bibinfo {title} {{The PyCBC
  search for gravitational waves from compact binary coalescence}},\ }\href
  {https://doi.org/10.1088/0264-9381/33/21/215004} {\bibfield  {journal}
  {\bibinfo  {journal} {Classical and Quantum Gravity}\ }\textbf {\bibinfo
  {volume} {33}},\ \bibinfo {eid} {215004} (\bibinfo {year} {2016})},\ \Eprint
  {https://arxiv.org/abs/1508.02357} {arXiv:1508.02357 [gr-qc]} \BibitemShut
  {NoStop}%
\bibitem [{\citenamefont {{Messick}}\ \emph {et~al.}(2017)\citenamefont
  {{Messick}} \emph {et~al.}}]{Messick2017}%
  \BibitemOpen
  \bibfield  {author} {\bibinfo {author} {\bibfnamefont {C.}~\bibnamefont
  {{Messick}}} \emph {et~al.},\ }\bibfield  {title} {\bibinfo {title}
  {{Analysis framework for the prompt discovery of compact binary mergers in
  gravitational-wave data}},\ }\href
  {https://doi.org/10.1103/PhysRevD.95.042001} {\bibfield  {journal} {\bibinfo
  {journal} {\prd}\ }\textbf {\bibinfo {volume} {95}},\ \bibinfo {eid} {042001}
  (\bibinfo {year} {2017})},\ \Eprint {https://arxiv.org/abs/1604.04324}
  {arXiv:1604.04324 [astro-ph.IM]} \BibitemShut {NoStop}%
\bibitem [{\citenamefont {{Chatziioannou}}\ \emph {et~al.}(2017)\citenamefont
  {{Chatziioannou}}, \citenamefont {{Clark}}, \citenamefont {{Bauswein}},
  \citenamefont {{Millhouse}}, \citenamefont {{Littenberg}},\ and\
  \citenamefont {{Cornish}}}]{Chatziioannou2017}%
  \BibitemOpen
  \bibfield  {author} {\bibinfo {author} {\bibfnamefont {K.}~\bibnamefont
  {{Chatziioannou}}}, \bibinfo {author} {\bibfnamefont {J.~A.}\ \bibnamefont
  {{Clark}}}, \bibinfo {author} {\bibfnamefont {A.}~\bibnamefont {{Bauswein}}},
  \bibinfo {author} {\bibfnamefont {M.}~\bibnamefont {{Millhouse}}}, \bibinfo
  {author} {\bibfnamefont {T.~B.}\ \bibnamefont {{Littenberg}}},\ and\ \bibinfo
  {author} {\bibfnamefont {N.}~\bibnamefont {{Cornish}}},\ }\bibfield  {title}
  {\bibinfo {title} {{Inferring the post-merger gravitational wave emission
  from binary neutron star coalescences}},\ }\href
  {https://doi.org/10.1103/PhysRevD.96.124035} {\bibfield  {journal} {\bibinfo
  {journal} {\prd}\ }\textbf {\bibinfo {volume} {96}},\ \bibinfo {eid} {124035}
  (\bibinfo {year} {2017})},\ \Eprint {https://arxiv.org/abs/1711.00040}
  {arXiv:1711.00040 [gr-qc]} \BibitemShut {NoStop}%
\bibitem [{\citenamefont {{Abbott}}\ \emph {et~al.}(2019)\citenamefont
  {{Abbott}} \emph {et~al.}}]{LVC2019}%
  \BibitemOpen
  \bibfield  {author} {\bibinfo {author} {\bibfnamefont {B.~P.}\ \bibnamefont
  {{Abbott}}} \emph {et~al.} (\bibinfo {collaboration} {{LIGO Scientific
  Collaboration} and {Virgo Collaboration}}),\ }\bibfield  {title} {\bibinfo
  {title} {{GWTC-1: A Gravitational-Wave Transient Catalog of Compact Binary
  Mergers Observed by LIGO and Virgo during the First and Second Observing
  Runs}},\ }\href {https://doi.org/10.1103/PhysRevX.9.031040} {\bibfield
  {journal} {\bibinfo  {journal} {Physical Review X}\ }\textbf {\bibinfo
  {volume} {9}},\ \bibinfo {eid} {031040} (\bibinfo {year} {2019})},\ \Eprint
  {https://arxiv.org/abs/1811.12907} {arXiv:1811.12907 [astro-ph.HE]}
  \BibitemShut {NoStop}%
\bibitem [{\citenamefont {{Ghonge}}\ \emph {et~al.}(2020)\citenamefont
  {{Ghonge}}, \citenamefont {{Chatziioannou}}, \citenamefont {{Clark}},
  \citenamefont {{Littenberg}}, \citenamefont {{Millhouse}}, \citenamefont
  {{Cadonati}},\ and\ \citenamefont {{Cornish}}}]{Ghonge2020}%
  \BibitemOpen
  \bibfield  {author} {\bibinfo {author} {\bibfnamefont {S.}~\bibnamefont
  {{Ghonge}}}, \bibinfo {author} {\bibfnamefont {K.}~\bibnamefont
  {{Chatziioannou}}}, \bibinfo {author} {\bibfnamefont {J.~A.}\ \bibnamefont
  {{Clark}}}, \bibinfo {author} {\bibfnamefont {T.}~\bibnamefont
  {{Littenberg}}}, \bibinfo {author} {\bibfnamefont {M.}~\bibnamefont
  {{Millhouse}}}, \bibinfo {author} {\bibfnamefont {L.}~\bibnamefont
  {{Cadonati}}},\ and\ \bibinfo {author} {\bibfnamefont {N.}~\bibnamefont
  {{Cornish}}},\ }\bibfield  {title} {\bibinfo {title} {{Reconstructing
  gravitational wave signals from binary black hole mergers with minimal
  assumptions}},\ }\href {https://doi.org/10.1103/PhysRevD.102.064056}
  {\bibfield  {journal} {\bibinfo  {journal} {\prd}\ }\textbf {\bibinfo
  {volume} {102}},\ \bibinfo {eid} {064056} (\bibinfo {year} {2020})},\ \Eprint
  {https://arxiv.org/abs/2003.09456} {arXiv:2003.09456 [gr-qc]} \BibitemShut
  {NoStop}%
\bibitem [{\citenamefont {{D{\'a}lya}}\ \emph {et~al.}(2021)\citenamefont
  {{D{\'a}lya}}, \citenamefont {{Raffai}},\ and\ \citenamefont
  {{B{\'e}csy}}}]{Dalya2021}%
  \BibitemOpen
  \bibfield  {author} {\bibinfo {author} {\bibfnamefont {G.}~\bibnamefont
  {{D{\'a}lya}}}, \bibinfo {author} {\bibfnamefont {P.}~\bibnamefont
  {{Raffai}}},\ and\ \bibinfo {author} {\bibfnamefont {B.}~\bibnamefont
  {{B{\'e}csy}}},\ }\bibfield  {title} {\bibinfo {title} {{Bayesian
  reconstruction of gravitational-wave signals from binary black holes with
  nonzero eccentricities}},\ }\href {https://doi.org/10.1088/1361-6382/abd7bf}
  {\bibfield  {journal} {\bibinfo  {journal} {Classical and Quantum Gravity}\
  }\textbf {\bibinfo {volume} {38}},\ \bibinfo {eid} {065002} (\bibinfo {year}
  {2021})},\ \Eprint {https://arxiv.org/abs/2006.06256} {arXiv:2006.06256
  [astro-ph.HE]} \BibitemShut {NoStop}%
\bibitem [{\citenamefont {{B{\'e}csy}}\ \emph {et~al.}(2017)\citenamefont
  {{B{\'e}csy}}, \citenamefont {{Raffai}}, \citenamefont {{Cornish}},
  \citenamefont {{Essick}}, \citenamefont {{Kanner}}, \citenamefont
  {{Katsavounidis}}, \citenamefont {{Littenberg}}, \citenamefont
  {{Millhouse}},\ and\ \citenamefont {{Vitale}}}]{Becsy2017}%
  \BibitemOpen
  \bibfield  {author} {\bibinfo {author} {\bibfnamefont {B.}~\bibnamefont
  {{B{\'e}csy}}}, \bibinfo {author} {\bibfnamefont {P.}~\bibnamefont
  {{Raffai}}}, \bibinfo {author} {\bibfnamefont {N.~J.}\ \bibnamefont
  {{Cornish}}}, \bibinfo {author} {\bibfnamefont {R.}~\bibnamefont {{Essick}}},
  \bibinfo {author} {\bibfnamefont {J.}~\bibnamefont {{Kanner}}}, \bibinfo
  {author} {\bibfnamefont {E.}~\bibnamefont {{Katsavounidis}}}, \bibinfo
  {author} {\bibfnamefont {T.~B.}\ \bibnamefont {{Littenberg}}}, \bibinfo
  {author} {\bibfnamefont {M.}~\bibnamefont {{Millhouse}}},\ and\ \bibinfo
  {author} {\bibfnamefont {S.}~\bibnamefont {{Vitale}}},\ }\bibfield  {title}
  {\bibinfo {title} {{Parameter Estimation for Gravitational-wave Bursts with
  the BayesWave Pipeline}},\ }\href {https://doi.org/10.3847/1538-4357/aa63ef}
  {\bibfield  {journal} {\bibinfo  {journal} {\apj}\ }\textbf {\bibinfo
  {volume} {839}},\ \bibinfo {eid} {15} (\bibinfo {year} {2017})},\ \Eprint
  {https://arxiv.org/abs/1612.02003} {arXiv:1612.02003 [astro-ph.HE]}
  \BibitemShut {NoStop}%
\bibitem [{\citenamefont {{Pankow}}\ \emph {et~al.}(2018)\citenamefont
  {{Pankow}} \emph {et~al.}}]{Pankow2018}%
  \BibitemOpen
  \bibfield  {author} {\bibinfo {author} {\bibfnamefont {C.}~\bibnamefont
  {{Pankow}}} \emph {et~al.},\ }\bibfield  {title} {\bibinfo {title}
  {{Mitigation of the instrumental noise transient in gravitational-wave data
  surrounding GW170817}},\ }\href {https://doi.org/10.1103/PhysRevD.98.084016}
  {\bibfield  {journal} {\bibinfo  {journal} {\prd}\ }\textbf {\bibinfo
  {volume} {98}},\ \bibinfo {eid} {084016} (\bibinfo {year} {2018})},\ \Eprint
  {https://arxiv.org/abs/1808.03619} {arXiv:1808.03619 [gr-qc]} \BibitemShut
  {NoStop}%
\bibitem [{\citenamefont {{Chatziioannou}}\ \emph {et~al.}(2021)\citenamefont
  {{Chatziioannou}}, \citenamefont {{Cornish}}, \citenamefont {{Wijngaarden}},\
  and\ \citenamefont {{Littenberg}}}]{Chatziioannou2021}%
  \BibitemOpen
  \bibfield  {author} {\bibinfo {author} {\bibfnamefont {K.}~\bibnamefont
  {{Chatziioannou}}}, \bibinfo {author} {\bibfnamefont {N.}~\bibnamefont
  {{Cornish}}}, \bibinfo {author} {\bibfnamefont {M.}~\bibnamefont
  {{Wijngaarden}}},\ and\ \bibinfo {author} {\bibfnamefont {T.~B.}\
  \bibnamefont {{Littenberg}}},\ }\bibfield  {title} {\bibinfo {title}
  {{Modeling compact binary signals and instrumental glitches in gravitational
  wave data}},\ }\href {https://doi.org/10.1103/PhysRevD.103.044013} {\bibfield
   {journal} {\bibinfo  {journal} {\prd}\ }\textbf {\bibinfo {volume} {103}},\
  \bibinfo {eid} {044013} (\bibinfo {year} {2021})},\ \Eprint
  {https://arxiv.org/abs/2101.01200} {arXiv:2101.01200 [gr-qc]} \BibitemShut
  {NoStop}%
\bibitem [{\citenamefont {{Gill}}\ \emph {et~al.}(2018)\citenamefont {{Gill}},
  \citenamefont {{Wang}}, \citenamefont {{Valdez}}, \citenamefont
  {{Szczepanczyk}}, \citenamefont {{Zanolin}},\ and\ \citenamefont
  {{Mukherjee}}}]{Gill2018}%
  \BibitemOpen
  \bibfield  {author} {\bibinfo {author} {\bibfnamefont {K.}~\bibnamefont
  {{Gill}}}, \bibinfo {author} {\bibfnamefont {W.}~\bibnamefont {{Wang}}},
  \bibinfo {author} {\bibfnamefont {O.}~\bibnamefont {{Valdez}}}, \bibinfo
  {author} {\bibfnamefont {M.}~\bibnamefont {{Szczepanczyk}}}, \bibinfo
  {author} {\bibfnamefont {M.}~\bibnamefont {{Zanolin}}},\ and\ \bibinfo
  {author} {\bibfnamefont {S.}~\bibnamefont {{Mukherjee}}},\ }\bibfield
  {title} {\bibinfo {title} {{Enhancing the Sensitivity of Searches for
  Gravitational Waves from Core-Collapse Supernovae with a Bayesian
  classification of candidate events}},\ }\href@noop {} {\bibfield  {journal}
  {\bibinfo  {journal} {arXiv e-prints}\ ,\ \bibinfo {eid} {arXiv:1802.07255}}
  (\bibinfo {year} {2018})},\ \Eprint {https://arxiv.org/abs/1802.07255}
  {arXiv:1802.07255 [astro-ph.HE]} \BibitemShut {NoStop}%
\bibitem [{\citenamefont {{Powell}}\ and\ \citenamefont
  {{M{\"u}ller}}(2020)}]{Powell2020}%
  \BibitemOpen
  \bibfield  {author} {\bibinfo {author} {\bibfnamefont {J.}~\bibnamefont
  {{Powell}}}\ and\ \bibinfo {author} {\bibfnamefont {B.}~\bibnamefont
  {{M{\"u}ller}}},\ }\bibfield  {title} {\bibinfo {title} {{Three-dimensional
  core-collapse supernova simulations of massive and rotating progenitors}},\
  }\href {https://doi.org/10.1093/mnras/staa1048} {\bibfield  {journal}
  {\bibinfo  {journal} {\mnras}\ }\textbf {\bibinfo {volume} {494}},\ \bibinfo
  {pages} {4665} (\bibinfo {year} {2020})},\ \Eprint
  {https://arxiv.org/abs/2002.10115} {arXiv:2002.10115 [astro-ph.HE]}
  \BibitemShut {NoStop}%
\bibitem [{\citenamefont {{Cautun}}\ \emph {et~al.}(2020)\citenamefont
  {{Cautun}}, \citenamefont {{Ben{\'\i}tez-Llambay}}, \citenamefont {{Deason}},
  \citenamefont {{Frenk}}, \citenamefont {{Fattahi}}, \citenamefont
  {{G{\'o}mez}}, \citenamefont {{Grand}}, \citenamefont {{Oman}}, \citenamefont
  {{Navarro}},\ and\ \citenamefont {{Simpson}}}]{Cautun2020}%
  \BibitemOpen
  \bibfield  {author} {\bibinfo {author} {\bibfnamefont {M.}~\bibnamefont
  {{Cautun}}}, \bibinfo {author} {\bibfnamefont {A.}~\bibnamefont
  {{Ben{\'\i}tez-Llambay}}}, \bibinfo {author} {\bibfnamefont {A.~J.}\
  \bibnamefont {{Deason}}}, \bibinfo {author} {\bibfnamefont {C.~S.}\
  \bibnamefont {{Frenk}}}, \bibinfo {author} {\bibfnamefont {A.}~\bibnamefont
  {{Fattahi}}}, \bibinfo {author} {\bibfnamefont {F.~A.}\ \bibnamefont
  {{G{\'o}mez}}}, \bibinfo {author} {\bibfnamefont {R.~J.~J.}\ \bibnamefont
  {{Grand}}}, \bibinfo {author} {\bibfnamefont {K.~A.}\ \bibnamefont {{Oman}}},
  \bibinfo {author} {\bibfnamefont {J.~F.}\ \bibnamefont {{Navarro}}},\ and\
  \bibinfo {author} {\bibfnamefont {C.~M.}\ \bibnamefont {{Simpson}}},\
  }\bibfield  {title} {\bibinfo {title} {{The milky way total mass profile as
  inferred from Gaia DR2}},\ }\href {https://doi.org/10.1093/mnras/staa1017}
  {\bibfield  {journal} {\bibinfo  {journal} {\mnras}\ }\textbf {\bibinfo
  {volume} {494}},\ \bibinfo {pages} {4291} (\bibinfo {year} {2020})},\ \Eprint
  {https://arxiv.org/abs/1911.04557} {arXiv:1911.04557 [astro-ph.GA]}
  \BibitemShut {NoStop}%
\bibitem [{\citenamefont {{Cornish}}\ \emph {et~al.}(2021)\citenamefont
  {{Cornish}}, \citenamefont {{Littenberg}}, \citenamefont {{B{\'e}csy}},
  \citenamefont {{Chatziioannou}}, \citenamefont {{Clark}}, \citenamefont
  {{Ghonge}},\ and\ \citenamefont {{Millhouse}}}]{Cornish2021}%
  \BibitemOpen
  \bibfield  {author} {\bibinfo {author} {\bibfnamefont {N.~J.}\ \bibnamefont
  {{Cornish}}}, \bibinfo {author} {\bibfnamefont {T.~B.}\ \bibnamefont
  {{Littenberg}}}, \bibinfo {author} {\bibfnamefont {B.}~\bibnamefont
  {{B{\'e}csy}}}, \bibinfo {author} {\bibfnamefont {K.}~\bibnamefont
  {{Chatziioannou}}}, \bibinfo {author} {\bibfnamefont {J.~A.}\ \bibnamefont
  {{Clark}}}, \bibinfo {author} {\bibfnamefont {S.}~\bibnamefont {{Ghonge}}},\
  and\ \bibinfo {author} {\bibfnamefont {M.}~\bibnamefont {{Millhouse}}},\
  }\bibfield  {title} {\bibinfo {title} {{BayesWave analysis pipeline in the
  era of gravitational wave observations}},\ }\href
  {https://doi.org/10.1103/PhysRevD.103.044006} {\bibfield  {journal} {\bibinfo
   {journal} {\prd}\ }\textbf {\bibinfo {volume} {103}},\ \bibinfo {eid}
  {044006} (\bibinfo {year} {2021})},\ \Eprint
  {https://arxiv.org/abs/2011.09494} {arXiv:2011.09494 [gr-qc]} \BibitemShut
  {NoStop}%
\bibitem [{\citenamefont {{Nakamura}}\ \emph {et~al.}(2016)\citenamefont
  {{Nakamura}}, \citenamefont {{Horiuchi}}, \citenamefont {{Tanaka}},
  \citenamefont {{Hayama}}, \citenamefont {{Takiwaki}},\ and\ \citenamefont
  {{Kotake}}}]{Nakamura2016}%
  \BibitemOpen
  \bibfield  {author} {\bibinfo {author} {\bibfnamefont {K.}~\bibnamefont
  {{Nakamura}}}, \bibinfo {author} {\bibfnamefont {S.}~\bibnamefont
  {{Horiuchi}}}, \bibinfo {author} {\bibfnamefont {M.}~\bibnamefont
  {{Tanaka}}}, \bibinfo {author} {\bibfnamefont {K.}~\bibnamefont {{Hayama}}},
  \bibinfo {author} {\bibfnamefont {T.}~\bibnamefont {{Takiwaki}}},\ and\
  \bibinfo {author} {\bibfnamefont {K.}~\bibnamefont {{Kotake}}},\ }\bibfield
  {title} {\bibinfo {title} {{Multimessenger signals of long-term core-collapse
  supernova simulations: synergetic observation strategies}},\ }\href
  {https://doi.org/10.1093/mnras/stw1453} {\bibfield  {journal} {\bibinfo
  {journal} {\mnras}\ }\textbf {\bibinfo {volume} {461}},\ \bibinfo {pages}
  {3296} (\bibinfo {year} {2016})},\ \Eprint {https://arxiv.org/abs/1602.03028}
  {arXiv:1602.03028 [astro-ph.HE]} \BibitemShut {NoStop}%
\bibitem [{\citenamefont {{Al Kharusi}}\ \emph {et~al.}(2021)\citenamefont {{Al
  Kharusi}} \emph {et~al.}}]{AlKharusi2021}%
  \BibitemOpen
  \bibfield  {author} {\bibinfo {author} {\bibfnamefont {S.}~\bibnamefont {{Al
  Kharusi}}} \emph {et~al.},\ }\bibfield  {title} {\bibinfo {title} {{SNEWS
  2.0: a next-generation supernova early warning system for multi-messenger
  astronomy}},\ }\href {https://doi.org/10.1088/1367-2630/abde33} {\bibfield
  {journal} {\bibinfo  {journal} {New Journal of Physics}\ }\textbf {\bibinfo
  {volume} {23}},\ \bibinfo {eid} {031201} (\bibinfo {year} {2021})},\ \Eprint
  {https://arxiv.org/abs/2011.00035} {arXiv:2011.00035 [astro-ph.HE]}
  \BibitemShut {NoStop}%
\bibitem [{\citenamefont {{Millhouse}}\ \emph {et~al.}(2018)\citenamefont
  {{Millhouse}}, \citenamefont {{Cornish}},\ and\ \citenamefont
  {{Littenberg}}}]{Millhouse2018}%
  \BibitemOpen
  \bibfield  {author} {\bibinfo {author} {\bibfnamefont {M.}~\bibnamefont
  {{Millhouse}}}, \bibinfo {author} {\bibfnamefont {N.~J.}\ \bibnamefont
  {{Cornish}}},\ and\ \bibinfo {author} {\bibfnamefont {T.}~\bibnamefont
  {{Littenberg}}},\ }\bibfield  {title} {\bibinfo {title} {{Bayesian
  reconstruction of gravitational wave bursts using chirplets}},\ }\href
  {https://doi.org/10.1103/PhysRevD.97.104057} {\bibfield  {journal} {\bibinfo
  {journal} {\prd}\ }\textbf {\bibinfo {volume} {97}},\ \bibinfo {eid} {104057}
  (\bibinfo {year} {2018})},\ \Eprint {https://arxiv.org/abs/1804.03239}
  {arXiv:1804.03239 [gr-qc]} \BibitemShut {NoStop}%
\bibitem [{\citenamefont {{Littenberg}}\ and\ \citenamefont
  {{Cornish}}(2015)}]{Littenberg2015}%
  \BibitemOpen
  \bibfield  {author} {\bibinfo {author} {\bibfnamefont {T.~B.}\ \bibnamefont
  {{Littenberg}}}\ and\ \bibinfo {author} {\bibfnamefont {N.~J.}\ \bibnamefont
  {{Cornish}}},\ }\bibfield  {title} {\bibinfo {title} {{Bayesian inference for
  spectral estimation of gravitational wave detector noise}},\ }\href
  {https://doi.org/10.1103/PhysRevD.91.084034} {\bibfield  {journal} {\bibinfo
  {journal} {\prd}\ }\textbf {\bibinfo {volume} {91}},\ \bibinfo {eid} {084034}
  (\bibinfo {year} {2015})},\ \Eprint {https://arxiv.org/abs/1410.3852}
  {arXiv:1410.3852 [gr-qc]} \BibitemShut {NoStop}%
\bibitem [{\citenamefont {{Brown}}(1991)}]{Brown1991}%
  \BibitemOpen
  \bibfield  {author} {\bibinfo {author} {\bibfnamefont {J.~C.}\ \bibnamefont
  {{Brown}}},\ }\bibfield  {title} {\bibinfo {title} {{Calculation of a
  constant Q spectral transform}},\ }\href {https://doi.org/10.1121/1.400476}
  {\bibfield  {journal} {\bibinfo  {journal} {Acoustical Society of America
  Journal}\ }\textbf {\bibinfo {volume} {89}},\ \bibinfo {pages} {425}
  (\bibinfo {year} {1991})}\BibitemShut {NoStop}%
\bibitem [{\citenamefont {{Kanner}}\ \emph {et~al.}(2016)\citenamefont
  {{Kanner}}, \citenamefont {{Littenberg}}, \citenamefont {{Cornish}},
  \citenamefont {{Millhouse}}, \citenamefont {{Xhakaj}}, \citenamefont
  {{Salemi}}, \citenamefont {{Drago}}, \citenamefont {{Vedovato}},\ and\
  \citenamefont {{Klimenko}}}]{Kanner2016}%
  \BibitemOpen
  \bibfield  {author} {\bibinfo {author} {\bibfnamefont {J.~B.}\ \bibnamefont
  {{Kanner}}}, \bibinfo {author} {\bibfnamefont {T.~B.}\ \bibnamefont
  {{Littenberg}}}, \bibinfo {author} {\bibfnamefont {N.}~\bibnamefont
  {{Cornish}}}, \bibinfo {author} {\bibfnamefont {M.}~\bibnamefont
  {{Millhouse}}}, \bibinfo {author} {\bibfnamefont {E.}~\bibnamefont
  {{Xhakaj}}}, \bibinfo {author} {\bibfnamefont {F.}~\bibnamefont {{Salemi}}},
  \bibinfo {author} {\bibfnamefont {M.}~\bibnamefont {{Drago}}}, \bibinfo
  {author} {\bibfnamefont {G.}~\bibnamefont {{Vedovato}}},\ and\ \bibinfo
  {author} {\bibfnamefont {S.}~\bibnamefont {{Klimenko}}},\ }\bibfield  {title}
  {\bibinfo {title} {{Leveraging waveform complexity for confident detection of
  gravitational waves}},\ }\href {https://doi.org/10.1103/PhysRevD.93.022002}
  {\bibfield  {journal} {\bibinfo  {journal} {\prd}\ }\textbf {\bibinfo
  {volume} {93}},\ \bibinfo {eid} {022002} (\bibinfo {year} {2016})},\ \Eprint
  {https://arxiv.org/abs/1509.06423} {arXiv:1509.06423 [astro-ph.IM]}
  \BibitemShut {NoStop}%
\bibitem [{\citenamefont {{Lee}}\ \emph {et~al.}(2021)\citenamefont {{Lee}},
  \citenamefont {{Millhouse}},\ and\ \citenamefont {{Melatos}}}]{Lee2021}%
  \BibitemOpen
  \bibfield  {author} {\bibinfo {author} {\bibfnamefont {Y.~S.~C.}\
  \bibnamefont {{Lee}}}, \bibinfo {author} {\bibfnamefont {M.}~\bibnamefont
  {{Millhouse}}},\ and\ \bibinfo {author} {\bibfnamefont {A.}~\bibnamefont
  {{Melatos}}},\ }\bibfield  {title} {\bibinfo {title} {{Enhancing the
  gravitational-wave burst detection confidence in expanded detector networks
  with the BayesWave pipeline}},\ }\href
  {https://doi.org/10.1103/PhysRevD.103.062002} {\bibfield  {journal} {\bibinfo
   {journal} {\prd}\ }\textbf {\bibinfo {volume} {103}},\ \bibinfo {eid}
  {062002} (\bibinfo {year} {2021})},\ \Eprint
  {https://arxiv.org/abs/2102.10816} {arXiv:2102.10816 [gr-qc]} \BibitemShut
  {NoStop}%
\bibitem [{\citenamefont {{Dolan}}\ \emph {et~al.}(2016)\citenamefont
  {{Dolan}}, \citenamefont {{Mathews}}, \citenamefont {{Lam}}, \citenamefont
  {{Quynh Lan}}, \citenamefont {{Herczeg}},\ and\ \citenamefont
  {{Dearborn}}}]{Dolan2016}%
  \BibitemOpen
  \bibfield  {author} {\bibinfo {author} {\bibfnamefont {M.~M.}\ \bibnamefont
  {{Dolan}}}, \bibinfo {author} {\bibfnamefont {G.~J.}\ \bibnamefont
  {{Mathews}}}, \bibinfo {author} {\bibfnamefont {D.~D.}\ \bibnamefont
  {{Lam}}}, \bibinfo {author} {\bibfnamefont {N.}~\bibnamefont {{Quynh Lan}}},
  \bibinfo {author} {\bibfnamefont {G.~J.}\ \bibnamefont {{Herczeg}}},\ and\
  \bibinfo {author} {\bibfnamefont {D.~S.~P.}\ \bibnamefont {{Dearborn}}},\
  }\bibfield  {title} {\bibinfo {title} {{Evolutionary Tracks for
  Betelgeuse}},\ }\href {https://doi.org/10.3847/0004-637X/819/1/7} {\bibfield
  {journal} {\bibinfo  {journal} {\apj}\ }\textbf {\bibinfo {volume} {819}},\
  \bibinfo {eid} {7} (\bibinfo {year} {2016})},\ \Eprint
  {https://arxiv.org/abs/1406.3143} {arXiv:1406.3143 [astro-ph.SR]}
  \BibitemShut {NoStop}%
\bibitem [{\citenamefont {{Pannarale}}\ \emph {et~al.}(2019)\citenamefont
  {{Pannarale}}, \citenamefont {{Macas}},\ and\ \citenamefont
  {{Sutton}}}]{Pannarale2019}%
  \BibitemOpen
  \bibfield  {author} {\bibinfo {author} {\bibfnamefont {F.}~\bibnamefont
  {{Pannarale}}}, \bibinfo {author} {\bibfnamefont {R.}~\bibnamefont
  {{Macas}}},\ and\ \bibinfo {author} {\bibfnamefont {P.~J.}\ \bibnamefont
  {{Sutton}}},\ }\bibfield  {title} {\bibinfo {title} {{Bayesian inference
  analysis of unmodelled gravitational-wave transients}},\ }\href
  {https://doi.org/10.1088/1361-6382/aaf76d} {\bibfield  {journal} {\bibinfo
  {journal} {Classical and Quantum Gravity}\ }\textbf {\bibinfo {volume}
  {36}},\ \bibinfo {eid} {035011} (\bibinfo {year} {2019})},\ \Eprint
  {https://arxiv.org/abs/1807.01939} {arXiv:1807.01939 [gr-qc]} \BibitemShut
  {NoStop}%
\end{thebibliography}%

\end{document}